\documentclass[letterpaper]{article}

\usepackage[T1]{fontenc}

\usepackage{geometry}
\usepackage{setspace}

\usepackage{achemso}

\usepackage{graphicx}
\usepackage{float}
\newfloat{scheme}{htbp}{los}
\floatname{scheme}{Scheme}
\floatname{chart}{Chart}
\newfloat{graph}{htbp}{loh}

\usepackage{chemformula} 
\usepackage[version = 4]{mhchem} 
\usepackage{dcolumn}
\usepackage{bm}
\usepackage{color}
\usepackage{amsmath}
\usepackage{booktabs} 
\usepackage{array} 
\usepackage{makecell}
\usepackage{colortbl}
\usepackage{xcolor} 
\usepackage[justification=centering,singlelinecheck=false]{caption}
\usepackage{braket}
\usepackage{subcaption}
\usepackage{caption}

\usepackage{authblk}
\author[1]{Yetmgeta Aklilu}
\affil[1]{ Department of Physics and Astronomy, Vanderbilt
University, Nashville,
Tennessee 37235, United States}
\author[2]{Matthew Shepherd}
\affil[2]{Department of Physics and Astronomy, Auburn University,
Auburn, Alabama 36830, United States}
\author[3]{Cody L. Covington}
\affil[3]{Department of Chemistry, Austin Peay State University,
Clarksville, Tennessee 37044, United States}
\author[1]{Kalman Varga}
\title{Quantum-electrodynamical time-dependent density functional
theory description of molecules in optical cavities}
\date{*Email:kalman.varga@vanderbilt.edu}

\begin{document}
\maketitle

\begin{abstract}
We introduce a quantum-electrodynamical time-dependent density
functional theory with a tensor-product representation (QED-TDDFT-TP)
to model molecules strongly coupled to quantized cavity fields. By
combining real-space electronic wavefunctions with truncated
Fock-space photon states, the method captures light–matter
correlations at a computational cost close to standard DFT. Benchmark
calculations show good agreement with QED-FCI and QED-CASCI for
ground-state energies and polaritonic spectra. Applications to weakly
bound dimers—including (H$_2$)$_2$, Ar$_2$, (H$_2$O)$_2$, and HF—demonstrate that
cavity confinement can significantly alter binding energies and
geometries in a polarization-dependent manner. The framework provides
an accurate and scalable tool for studying cavity-modified molecular
structure and interactions.
\end{abstract}

\section{Introduction}
Cavity quantum electrodynamics (cavity QED) is a fundamental framework
in quantum physics that studies the interaction between light and
matter in confined electromagnetic environments. Its importance spans
both theoretical understanding and practical applications.
Cavity QED provides precise control over light{--}matter interactions by
confining photons in optical cavities alongside atoms or other quantum
emitters. This confinement modifies the electromagnetic environment,
leading to phenomena {such as} the Purcell effect
\cite{PhysRevResearch.6.L012050}, where spontaneous
emission rates are enhanced or suppressed depending on the cavity
properties. These controlled interactions reveal fundamental aspects
of quantum mechanics, including entanglement between light and matter
\cite{PhysRevB.110.184306,welman2025lightmatterentanglementrealtimenuclearelectronic,Luo2024,PhysRevResearch.7.L012058}
and quantum superposition states \cite{Basov2025,doi:10.1021/acs.jpclett.4c03439}.
Light{--}matter coupling as a means to tune physical and chemical
properties has become a major focus of experimental research
\cite{https://doi.org/10.1002/anie.201107033,
Balili1007,PhysRevLett.114.196403,Xiang665,
doi:10.1021/acsphotonics.0c01224,Coles2014,Kasprzak2006,PhysRevLett.106.196405,Plumhof2014,
https://doi.org/10.1002/adma.201203682,Wang2021,BasovAsenjoGarciaSchuckZhuRubio+2021+549+577,PhysRevLett.125.123602,PhysRevX.5.041022}.
Theoretical investigations {have} also developed alongside the experimental
progress \cite{PhysRevLett.114.196402,PhysRevLett.121.253001,
PhysRevLett.128.156402,Riso2022,
https://doi.org/10.1002/qua.26750,
PhysRevLett.122.017401,DiStefano2019,PhysRevX.5.041022,
PhysRevLett.116.238301,Galego2016,Shalabney2015,Schafer4883,
Ruggenthaler2018,Flick15285,Flick3026,PhysRevLett.123.083201,Mandal,
FlickRiveraNarang,PhysRevLett.121.113002,Garcia-Vidaleabd0336,
Thomas615,PhysRevResearch.2.023262,doi:10.1063/5.0036283,doi:10.1063/5.0038748,doi:10.1063/5.0039256,
doi:10.1063/5.0012723,doi:10.1063/5.0021033,acs.jpcb.0c03227,PhysRevLett.119.136001,
doi:10.1063/5.0012723,Flick15285,doi:10.1021/acsphotonics.7b01279,PhysRevA.98.043801,
doi:10.1021/acs.jpclett.0c01556,doi:10.1021/acs.jctc.0c00618,doi:10.1021/acs.jpclett.0c03436,
doi:10.1021/acs.jctc.0c00469,PhysRevB.98.235123,PhysRevLett.122.193603,
Szidarovszky_2020,doi:10.1021/acs.jpclett.1c01570,doi:10.1021/jacs.2c00921,
doi:10.1063/5.0095552,doi:10.1021/acs.jpclett.1c02659,doi:10.1021/jacs.1c13201,
Cederbaum2021}.
Several excellent review articles survey the current state of
experimental and theoretical approaches to cavity light{--}matter
interactions. These encompass reviews on hybrid light{--}matter states
\cite{doi:10.1021/acs.accounts.6b00295,doi:10.1063/PT.3.4749,10.1063/5.0225932,D3CP01415K},
{\textit{ab initio}} computational methods
\cite{Ruggenthaler2018,doi:10.1063/5.0094956,doi:10.1021/acs.chemrev.2c00788}
, and
molecular polaritonics
\cite{doi:10.1146/annurev-physchem-090519-042621,doi:10.1021/acsphotonics.2c00048,
doi:10.1021/acsphotonics.1c01749}.
Describing coupled light{--}matter systems theoretically and
computationally presents significant challenges. The quantum many-body
problem involving electron{--}nuclear interactions is already complex, and
incorporating photon degrees of freedom makes it even more demanding.
Recent years have seen numerous approaches developed
\cite{PhysRevLett.127.273601,10.1063/5.0066427,doi:10.1063/5.0078795,doi:10.1063/5.0038748,
PhysRevX.10.041043,PhysRevLett.127.273601,doi:10.1063/5.0039256,doi:10.1021/acsphotonics.7b01279,
PhysRevA.98.043801,PhysRevA.90.012508,PhysRevLett.110.233001,doi:10.1063/5.0021033,doi:10.1063/5.0057542,10.1063/5.0123909,
Latini2019,10.1063/5.0233717,10.1063/5.0230565,10.1063/5.0188471}
that extend beyond the basic two-level atom model \cite{Jaynes1962ComparisonOQ}.
These methods typically build upon established many-body quantum techniques, adapting them to
account for photon interactions.

The Pauli{--}Fierz (PF) nonrelativistic quantum electrodynamics
Hamiltonian has emerged as the most practical framework
\cite{Ruggenthaler2018,Rokaj_2018,Mandal,acs.jpcb.0c03227,PhysRevB.98.235123}
for computational applications.
The Pauli{--}Fierz Hamiltonian is the fundamental theoretical framework
for describing nonrelativistic quantum electrodynamics (QED),
originally developed by Wolfgang Pauli and Markus Fierz in {1938}
\cite{doi:10.1098/rspa.1936.0111,Fierz,doi:10.1098/rspa.1939.0140}.
The PF Hamiltonian consists of three
components: the electronic Hamiltonian, the photonic Hamiltonian, and
an interaction term that couples electrons and photons. The presence
of this coupling term necessitates the use of a combined
electron{--}photon wave function, where electronic states are represented
through an appropriate basis set while photonic states are expressed
using the Fock-space representation. The Pauli{--}Fierz Hamiltonian is
characterized by several distinctive features. First, it operates
within a nonrelativistic framework, which differs from complete
relativistic QED by treating matter particles nonrelativistically
while preserving the quantum nature of the electromagnetic field.
Second, it employs minimal coupling \cite{PhysRevB.101.121104,PhysRevB.111.085114}
to describe light{--}matter
interactions, achieved by substituting the canonical momentum
$\mathbf{p}$ with $\mathbf{p}
{-\, e\,\mathbf{A}}$, where $\mathbf{A}$ represents the electromagnetic vector potential. Third,
many practical implementations, especially in cavity QED and molecular
physics, utilize the long-wavelength or dipole approximation
\cite{Maurer_2021}, which considerably reduces computational complexity.

Similar to the Schr\"odinger equation, the Pauli{--}Fierz Hamiltonian lacks
analytical solutions for multi-electron atoms. For a single-electron
atom or ion, the problem becomes tractable by constructing a product basis
from hydrogenic eigenfunctions and Fock basis states, allowing exact
diagonalization to yield the solution \cite{PhysRevA.110.043119}. However,
systems with more than one electron require numerical methods for their solution.

As is typical in electronic structure calculations, methodologies can
be broadly categorized into two distinct families: wave function{--}based
methods and density{--}based approaches. Wave function{--}based methods
\cite{PhysRevLett.127.273601,10.1063/5.0066427,doi:10.1063/5.0078795,doi:10.1063/5.0038748,PhysRevX.10.041043}
characteristically employ coupled electron{--}photon wave functions, and
their product structure leads to a substantial increase in
computational dimensionality. The coupled electron{--}photon wave function
can be written as
\begin{equation}
{\ket{\Psi} = \sum_{n,m} C_{nm}\,\Phi_{nm}\,\ket{n},}
\end{equation}
where $\Phi_{nm}$ is {a} many-body basis function representing the
electrons {(and nuclei, if present)}, $\vert n\rangle$ is a
Fock-space basis for
photons{,} and $C_{nm}$ are {linear} expansion
{coefficients}. The Fock-space
basis can represent a single {mode or multiple} photon modes. The $\Phi_{nm}$
notation emphasizes that the spatial basis functions can be different
for different photon {sectors} if needed.

The simplest approach, the cavity QED Hartree{--}Fock, extends the traditional Hartree{--}Fock method to
include quantized electromagnetic field modes within optical cavities.
This approach treats the coupled electron{--}photon system using a
mean-field approximation, where electrons experience an effective
field created by all other electrons and the cavity photon modes.
The method employs a polaritonic wave function ansatz that is
typically written as a product of an electronic Slater determinant and
a photon state. Refs.
\cite{doi:10.1063/5.0078795,doi:10.1063/5.0038748,PhysRevX.10.041043,
PhysRevResearch.2.023262,mordovina2020polaritonic}
employ a coupled-cluster
(CC) methodology that constructs a reference wave function from the
direct product of a Hartree{--}Fock Slater determinant and the photon
vacuum state. The ground-state QED-CC wave function is then defined by
applying an exponentiated cluster operator to this product state. The
primary advantage of this method lies in its systematic improvability.
The traditional Complete Active Space Configuration Interaction (CASCI) approach
has been also extended to include quantized electromagnetic
field modes \cite{vu_cavity_2024}. In the CASCI ansatz for the electronic subspace, a subset
of active electrons and orbitals are identified, where a full CI
expansion is performed within that active space.
In QED-CASCI, this framework is generalized to
simultaneously treat both electronic correlation within the active
space and the coupling to cavity photon modes.

The stochastic variational method (QED-SVM)
\cite{PhysRevLett.127.273601,10.1063/5.0257034,10.1063/5.0066427,suzuki1998stochastic}
similarly employs a
product form combining matter and photonic wave functions, but differs
in its treatment of the matter component through explicitly correlated
Gaussian basis states. Variational parameters are optimized via
stochastic selection procedures, yielding highly precise energies and
wave functions. Due to the $N!$ scaling {of} explicit antisymmetrization of the
$N$-particle basis functions{,} the practical application of the QED-SVM approach is restricted to small
atomic and molecular systems.

The density-based approach, namely {quantum electrodynamical density functional theory (QED-DFT)}, is an
extension of traditional density functional theory (DFT)
\cite{Tokatly2013,Flick2015,flick_atoms_2017,Ruggenthaler2018,flick2019light,Ruggenthaler2014,
Flick2015,ruggenthaler2018quantum,Pellegrini2015}.
QED-DFT bridges the gap between quantum optics and
electronic-structure theory, making it possible to describe phenomena
where light and matter interact strongly, such as in optical cavities.
{QED-DFT} is an exact
reformulation of the PF Hamiltonian, based {on} many-body wave
theory. In QED-DFT, the complex coupled electron{--}photon system is represented
by two uncoupled, yet nonlinear, auxiliary quantum systems. The
electrons are described by the usual DFT equation which now contains
potentials describing the interaction of light and matter. A separate
Maxwell-like equation is used for the photons.
The QED-DFT calculations mostly use real-space bases but extensions to Gaussian
basis representation also exist \cite{yang2021quantum,doi:10.1021/acs.jpclett.3c01294}.
{Combinations} of QED-DFT with macroscopic QED
\cite{Svendsen2021,doi:10.1021/acs.jctc.3c00967}, {and} extension{s} to Dicke
\cite{Bakkestuen2025,PhysRevB.108.235424} and Rabi models
\cite{doi:10.1021/acs.jpca.4c07690} have also been developed. 

Standard electronic exchange-correlation (XC) functionals are
inadequate for QED-DFT because they fail to account for
electron-photon correlations that emerge under strong light-matter
coupling. This limitation results in inaccurate predictions of
polaritonic energy levels, ground-state modifications, and photon
distribution statistics, making the creation of specialized QED
exchange-correlation (QED-XC) potentials crucial for these systems
\cite{Ruggenthaler2014,Tokatly2013,Flick2015}.
Recent theoretical advances—including optimized-effective-potential
formulations and exact-model benchmarks—have established the formal
structure of QED-XC and demonstrated the role of photon-mediated
electron–electron interactions in modifying electronic structure and
dynamics \cite{Pellegrini2015,Flick2017,Buchholz2019,
PhysRevLett.134.073002}. These
developments enable ab initio predictions of cavity-induced
modifications to excitons, charge transfer, and chemical reactivity,
successfully reproducing observed Rabi splittings and vacuum-field
effects in molecules and materials \cite{Latini2019,Hirai2020,Sentef2020}.

Our QED-DFT methodology employs a coupled electron{--}photon wave
function analogous to those utilized in wave function{--}based methods.
This wave function is constructed on a tensor product combining a
spatial grid with a Fock-state representation. To differentiate this
framework from previously discussed QED-DFT approaches, we designate
the current method as QED-DFT-TP and QED-TDDFT-TP. The QED-DFT-TP
represents a specific implementation of QED-DFT {that adopts}
an
alternative ansatz through the use of a coupled electron{--}photon wave
function. While the tensor product formulation elevates the computational
dimensionality, it maintains the discrete nature of quantized photon
states. The coupled electron{--}photon wave function offers
an enhanced characterization of light{--}matter interactions through the
calculation of spatial wave functions within individual photon sectors.
In this approach, each molecular orbital is paired with distinct Fock
basis states representing quantized photon modes. The light{--}matter
interaction component of the Hamiltonian governs the coupling between
orbital elements across various photon states. The orthogonality of
Fock states maintains {the} sparse structure
characteristic of real-space DFT Hamiltonians. This sparsity enables
the implementation of computationally efficient iterative
diagonalization techniques commonly employed in conventional
real-space DFT methodologies.

This paper aims to use the QED-DFT-TP methodology for computing
various physical properties of molecules within optical cavities,
investigate the influence of cavity parameters on these properties,
and benchmark the results against established theoretical approaches.
Small molecules, including LiH, BH{$_3$}, {Ar$_2$}, H{$_2$}, HF and water dimers
will be used as examples.

\section{Formalism}
The systems we consider in this paper are all nonrelativistic and as a 
result the light{--}matter coupling can be consistently
described by the Pauli{--}Fierz nonrelativistic QED Hamiltonian
\cite{rokaj2018light, ruggenthaler2018quantum,tokatly2018conserving}. In addition, since we are working with small-sized systems,
we assume that the spatial variation of the cavity field is negligible over the dimension of the system,
i.e.\ we will use the dipole approximation. The PF Hamiltonian in the
velocity gauge can be written as a sum of the kinetic energy,
Kohn{--}Sham potential{,} and the photonic Hamiltonian,
\begin{equation}
H_V={\frac{1}{2m}\left(i\hbar \nabla+e
\hat{\mathbf{A}}\right)^2} + V_{KS}(\mathbf{r})+
\sum_{{\alpha=1}}^{{N_p}}{\frac{1}{2}}\left[{p_{\alpha}^2+
\omega_{\alpha}^2 q_{\alpha}^2}\right],
\label{HV}
\end{equation}
where $V_{KS}(\mathbf{r})$ refers to the Kohn{--}Sham (KS) noninteracting potential adapted from {the} KS-TDDFT scheme \cite{runge1984density}{ }and is given by
\begin{equation}
V_{{KS}}(\mathbf{r})=V_{\mathrm{H}}{[\rho (\mathbf{r})]}+V_{\mathrm{XC}}{[\rho (\mathbf{r})]}
+V_{\mathrm{ion}}(\mathbf{r}),
\end{equation}
where $\rho$ is the electron density, $V_\mathrm{H}$ is the Hartree
potential, {$V_\mathrm{XC}$} is the exchange{--}correlation potential, and
$V_{\mathrm{ion}}$ is the external potential due to the ions. The
exchange{--}correlation potential {$V_{\mathrm{XC}}$} is
approximated using the generalized gradient approximation (GGA),
developed by Perdew et al. \cite{92PRB_GGA}.

In the long-wavelength limit, the vector potential is spatially
uniform over the matter extent,
\begin{equation}
\hat{\mathbf A} = \sum_\alpha {\mathcal A_\alpha}\,\boldsymbol{\varepsilon}{_\alpha}\,\hat q_\alpha,
\qquad
{\mathcal A_\alpha} \equiv \sqrt{\frac{\hbar}{{\varepsilon_0}
V\,\omega_\alpha}},
\label{vecpot}
\end{equation}
with polarization $\boldsymbol{\varepsilon}_\alpha$, quantization
volume $V$, and frequency $\omega_\alpha$.
The expansion of the kinetic term in \eqref{HV} contains the
paramagnetic coupling {$\frac{e}{m}\hat{\mathbf p}\!\cdot\!\hat{\mathbf A}$}
and the diamagnetic (seagull) term {$\frac{e^2}{2m}\hat{\mathbf A}^2$}.
By introducing
\begin{equation}
\boldsymbol{\lambda}_{{\alpha}}=\frac{\boldsymbol{\varepsilon}{_\alpha}}{\sqrt{{\varepsilon_0}
V}},
\end{equation}
the {para}magnetic term becomes
\begin{equation}
\frac{e}{m}\sum_\alpha
\sqrt{\frac{\hbar}{\omega_\alpha}} \hat{\mathbf
p}\!\cdot\!\boldsymbol{\lambda}{_\alpha}\hat{q}{_\alpha}
\end{equation}
and the diamagnetic term
\begin{equation}
\frac{e^2}{2m}\sum_\alpha
\frac{\hbar}{\omega_\alpha}
\boldsymbol{\lambda}{_\alpha}^2\hat{q}{_\alpha}^2.
\end{equation}
{T}he paramagnetic interaction links photon states that differ by one
quantum number ($\Delta {n}=\pm 1$)
, while the diamagnetic interaction connects photon states with
quantum number changes of $\Delta n=0,\pm 2$.
In the special case where the diamagnetic term couples photon states
with identical quantum numbers ($\Delta n=0$), it behaves analogously
to the dipole self-interaction (DSI)
found in the length-gauge formulation, introduced below.

The Hamiltonian can also be transformed into {the} length gauge (see
Appendix \ref{appa}):
\begin{eqnarray}
\hat H{_L} &=& 
{\frac{1}{2m}\left(i\hbar \nabla+e
\hat{\mathbf{A}}\right)^2} + V_{KS}(\mathbf{r})+
{+\;} \sum_\alpha \frac{\hbar\omega{_\alpha}}{2}\bigl(\hat q_\alpha^2 + \hat
p_\alpha^2\bigr) \nonumber\\
&\;-\;& \sum_\alpha {\hat{\mathbf
D}\!\cdot\!
\hat{\mathbf E}_\alpha}
\;+\; \sum_\alpha
\bigl(\boldsymbol{\lambda}_{\alpha}
\!\cdot\!\hat{\mathbf D}\bigr)^2,
\label{HL}
\end{eqnarray}
where $\mathbf{D}$ is the dipole moment { and $\hat{\mathbf E}_\alpha$ the transverse electric field of mode $\alpha$}.
The first term of the second line couples photon states with
$\Delta n=\pm 1${.} The last term of the second line is the DSI,
coupling only states with $\Delta n=0$. This term
is always present and, contrary to what was believed previously, plays a crucial {role}
in the variational
formulation of the eigenvalue problem\cite{rokaj2018light}. The length
and the velocity-gauge Hamiltonians give identical results as
discussed in Appendix \ref{appa}.

The coupled system is described by orbitals defined on a tensor product of a real-space and a Fock-space. At the KS level, we can represent the orbitals as
\begin{equation}
\Phi{_{mn}}=\phi{_{mn}}(\mathbf{r}) {\ket{n}},
\ \ \ \ \ (m=1,{\ldots},N_{{occ}}),
\ \ \ \ \ (n=0,{\ldots},N_{{F}}),
\end{equation}
where {$\ket{n}$} is the Fock-space basis for the photons, $N_{{F}}$ is
the dimension of the Fock-space, and $N_{{occ}}$ is the number of
orbitals. For this paper, we will assume that there is one dominant
mode and we can ignore all others, i.e.\ {$N_p=1$}. Our system is, thus, described by a {four-}dimensional ({4D}) grid: {$N_x\times N_y \times N_z\times N_F$}, where $N_x,N_y,N_z$ are the number of grid points in {Cartesian} real space and $N_F$ refers to the size of the truncated Fock-space, i.e.\ 
the vacuum state {$\ket{0}$} has {$N_F=1$}.
Because the Fock-basis states are orthogonal, the elements of the overlap matrix are given by
\begin{equation}
\left(\Phi{_{mn}}\vert\Phi{_{m'n'}}\right)=
\langle\phi{_{mn}}\vert\phi{_{m'n}}\rangle \delta_{nn'},
\end{equation}
where the round bracket stands for integration over both real and
Fock-space and the angle bracket is integration over only {the} real part,
\begin{equation}
\langle\phi{_{mn}}\vert\phi{_{m'n}}\rangle =\sum_{i j k}
\phi{_{mn}}({x_i,y_j,z_k}) \phi{_{m'n}}({x_i,y_j,z_k}).
\end{equation}
The calculation of the matrix elements can be simplified further by
orthogonalizing the real part of the orbitals for each Fock-state using the Gram{--}Schmidt method. This new orthogonal set can then be normalized
\begin{equation}
\sum{_{n=0}^{N_F}}\vert\hat{\phi}_{mn}\vert^2=1.
\end{equation}
where $\hat{\phi}{_{mn}}$ with $(m=0,{\ldots},N{_{occ}})$ represents the orthogonalized set of basis for the same photon state {$\ket{n}$}.
In the present work, the minimization for the coupled light{--}matter orbitals is carried out by the conjugate-gradient {method}. The construction of the Hamiltonian matrix in the coupled basis is described in detail in our previous work 
\cite{malave2022real}.

The ground-state calculation follows conventional DFT approaches using
steepest descent or conjugate gradient approaches to calculate the
energies and the orbitals. 

The ground state orbitals will be used to initialize the time
propagation
\begin{equation}
\hat{\Phi}_{m}(\mathbf{r},t=0)=
\hat{\Phi}_{m}(\mathbf{r}).
\end{equation}
Any time propagation method typically used for TDDFT can be used here,
and in this work
we use Taylor time propagation \cite{yabana1996time}
\begin{equation}
\hat{\Phi}_{m}(\mathbf{r},t+\Delta t)={\rm e}^{-iH\Delta t}
\hat{\Phi}_{m}(\mathbf{r},t)=\sum_{j=0}^4 {(-i\Delta t)^j\over j!}
H^j\hat{\Phi}_{m}(\mathbf{r},t),
\end{equation}
where $\Delta t$ is chosen to be sufficiently small to conserve the
norm of the orbitals
during propagation. 

To compute the absorption spectrum we apply a weak instantaneous
perturbation (a
``delta kick'') to the ground-state wave function,
\begin{equation}
\Phi_m^{+} = e^{i\kappa \hat{x}}\Phi_m,
\end{equation}
where $\kappa$ is small and $\hat{x}$ is the dipole operator. The
time-dependent
Schr\"odinger equation is then propagated,
and the dipole moment $d(t)=\sum_m\langle \Phi_m(t)|\hat{x}|\Phi_m(t)\rangle$ is
recorded.
The absorption spectrum is obtained from the imaginary part of the
Fourier
transform of the dipole response,
\begin{equation}
S(\omega) \propto \omega\, \mathrm{Im}\!\left[ \int_0^{T} e^{i\omega
t} d(t)\, dt \right].
\end{equation}

\section{Results}
The primary objective of these calculations is to benchmark the
QED-DFT-TP approach against other established methods across various
molecules and molecular dimers. This comparison is particularly timely
given the rapid development of numerous new approaches in this field.
To ensure a correct comparison, we have employed identical parameters
across all methods, including coupling strengths, cavity frequencies,
and molecular geometries.
We represent $\boldsymbol{\lambda}=\lambda \boldsymbol{\varepsilon}$ where
$\boldsymbol{\varepsilon}$
is a unit vector describing the polarization of the cavity mode, e.g.
{$(1,0,0)$,} and $\lambda$ is the coupling strength. In this work, we will
use $\lambda \le 0.1$, which, according to
{$\lambda=1/\sqrt{\varepsilon_0 V_{\mathrm{eff}}}$}, {corresponds to sub-nm$^3$ effective}
volumes \cite{carnegie2018room,benz2016single}.
This range coincides with volumes achieved in picocavity experiments
\cite{carnegie2018room,benz2016single}. 
The velocity-gauge Hamiltonian is employed in all calculations. The
differences of results in velocity and lengths gauges
are discussed  in Appendix A.
Since this calculation employs a finite-difference grid to represent
the wave functions, the computed total energies are sensitive to the
alignment between grid points and ionic positions. To maintain
consistency when investigating energy as a function of intermolecular
distance, we preserve the relative positioning by ensuring that
molecular coordinates remain commensurate with the underlying
computational lattice. This constraint limits our ability to position
molecules at arbitrary locations; instead, molecular displacements are
restricted to shifts with integer multiples of the grid spacing.

\subsection{Selected test cases}
This work examines the following 6 molecular systems under cavity
quantum electrodynamics conditions.
We begin by revisiting the potential energy surface (PES) of LiH,
which was previously explored in our earlier study
\cite{malave2022real}. Recent developments \cite{vu2024cavity} now
provide additional benchmarks for comparison with our methodology.
The second system of interest is BH$_3$, for which we calculate the Rabi splitting via time propagation of
molecular orbitals to obtain the absorption spectrum, enabling
comparison with the recent work of Ref.~\cite{vu2024cavity}.
Subsequently, we investigate the PES of an H$_2$
 dimer to characterize cavity-induced modifications to intermolecular
forces, with results benchmarked against those reported in Ref.
\cite{haugland2021intermolecular}. A parallel study of the Ar dimer
allows us to compare our approach with QED-DFT calculations from
Ref.\cite{PhysRevLett.134.073002}, providing a valuable
cross-validation between distinct QED-DFT formulations.
We then examine cavity effects on hydrogen bonding by computing the
PES of a water dimer, comparing our findings with
Ref.~\cite{vu2024cavity}.
Finally, we explore the orientational dependence of the
cavity-modified PES for an HF dimer, a system not previously
characterized in the literature.

\subsection{LiH}
\begin{figure}
\centering
\begin{subfigure}{0.45\textwidth}
 \includegraphics[width=\textwidth]{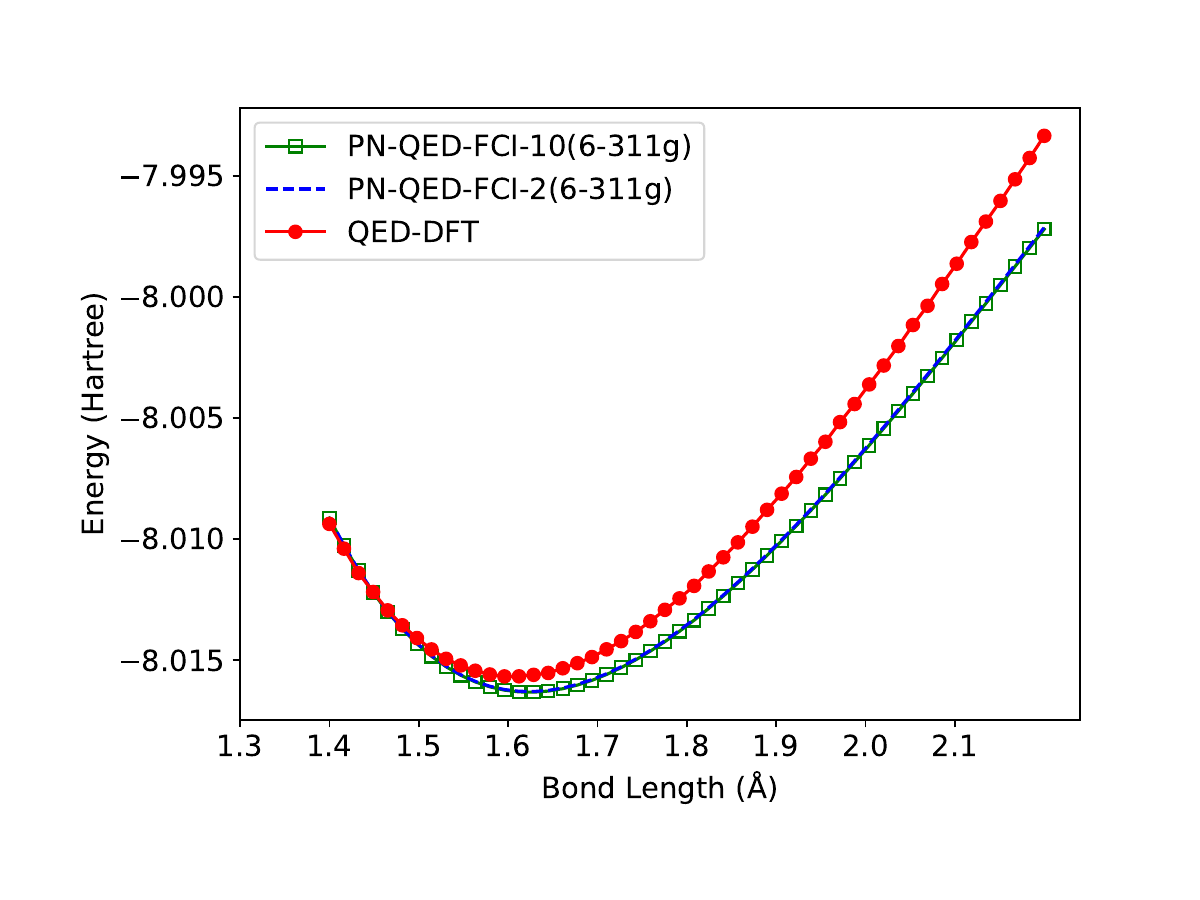}
\end{subfigure}
\hfill
\begin{subfigure}{0.45\textwidth}
 \includegraphics[width=\textwidth]{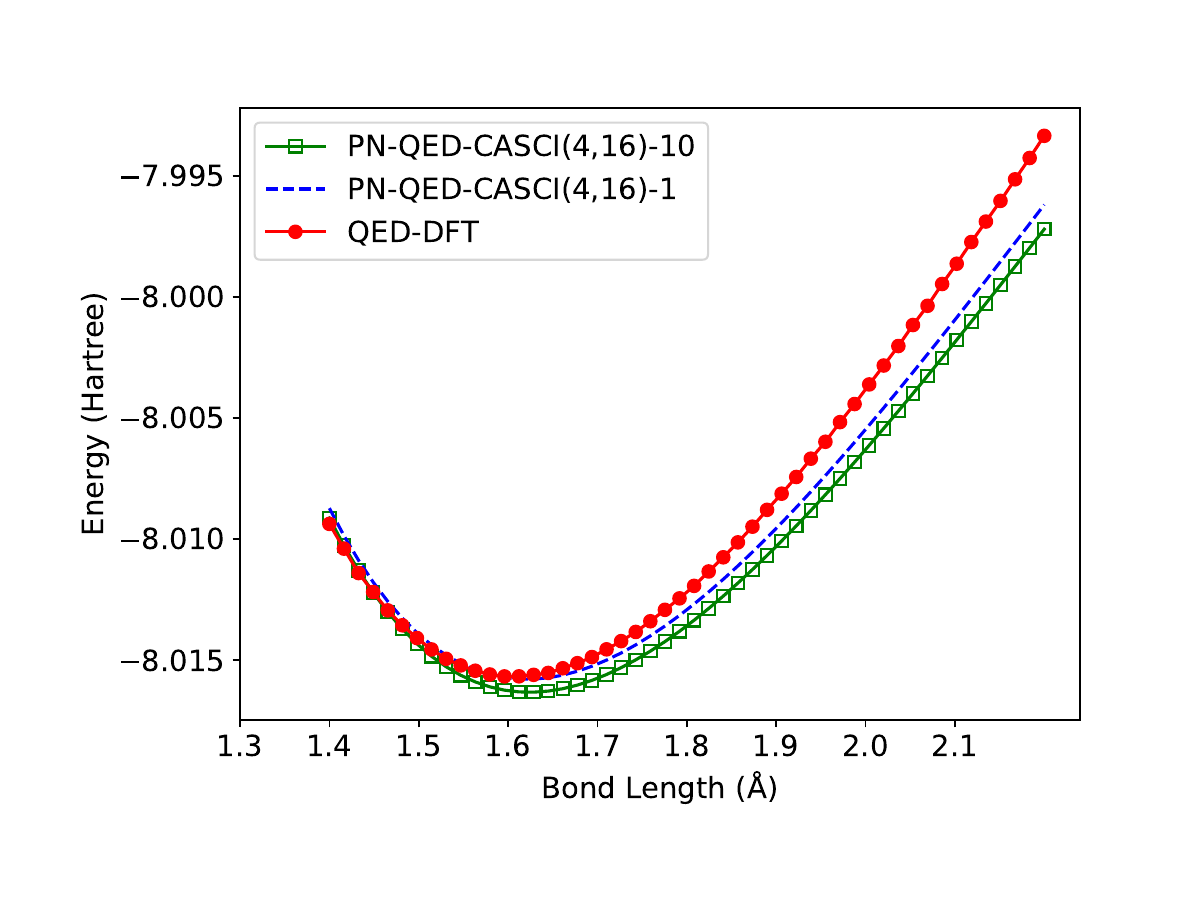}
\end{subfigure}
\caption{Comparison of PES calculated using QED-DFT-TP with PN-QED-FCI and PN-QED-CASCI. All graphs
use $\lambda=0.05$ and $\omega=0.121$. The left graph uses $N_F=1$,
while the right compares PN-QED-CASCI  $N_F=1$ and $N_F=10$ to
QED-DFT-TP.}
\label{fig:lih}
\end{figure}

In this section, we model the ground-state potential energy surface of
LiH and compare our results with Photon-Number Quantum
Electrodynamical Complete Active Space Configuration Interaction
(PN-QED-CASCI) and Photon-Number Quantum Electrodynamical Full
Configuration Interaction (PN-QED-FCI) calculations from
Ref.~\cite{vu2024cavity}.
The LiH molecule is oriented with its internuclear axis parallel to
the polarization vector of a cavity mode with frequency
$\omega=0.121$~a.u. This frequency is chosen to be resonant with the
molecule's lowest singlet excitation \cite{vu2024cavity}.

Fig.~\ref{fig:lih} presents a comparison of PN-QED-FCI,
PN-QED-CASCI, and QED-DFT-TP results for photon Fock spaces truncated
at $N_F=1$ and $N_F=10$. For $N_F=1$, only the $\ket{0}
$ Fock state is coupled to the electronic degrees of freedom, such
that exclusively the diamagnetic self-energy term contributes to the
cavity modification of the PES. To facilitate visual comparison, the
QED-DFT-TP energies have been uniformly shifted by $\Delta =
-7.264$ Hartree. All computational parameters were selected to match
those employed in the PN-QED-CASCI and PN-QED-FCI benchmarks of
Ref.\cite{vu2024cavity}.
The agreement between methodologies is excellent, with nearly
identical predictions for the equilibrium bond length. The primary
discrepancy emerges in the dissociation limit, which may stem from
either the compact 6-311G basis set employed in the wave function
methods or limitations of the GGA exchange-correlation functional in
QED-DFT-TP.
Comparison of the upper and lower panels of Fig.~\ref{fig:lih} reveals
rapid convergence with respect to photon space dimension for both wave
function and QED-DFT-TP approaches. Only the lowest few Fock states
exhibit significant coupling to the electronic subsystem, with the
$\ket{0}$ state accounting for approximately 98\% of the total photonic
probability.
\begin{figure*}[t]
 \centering
 \begin{subfigure}[t]{0.48\textwidth}
\includegraphics[width=\linewidth]{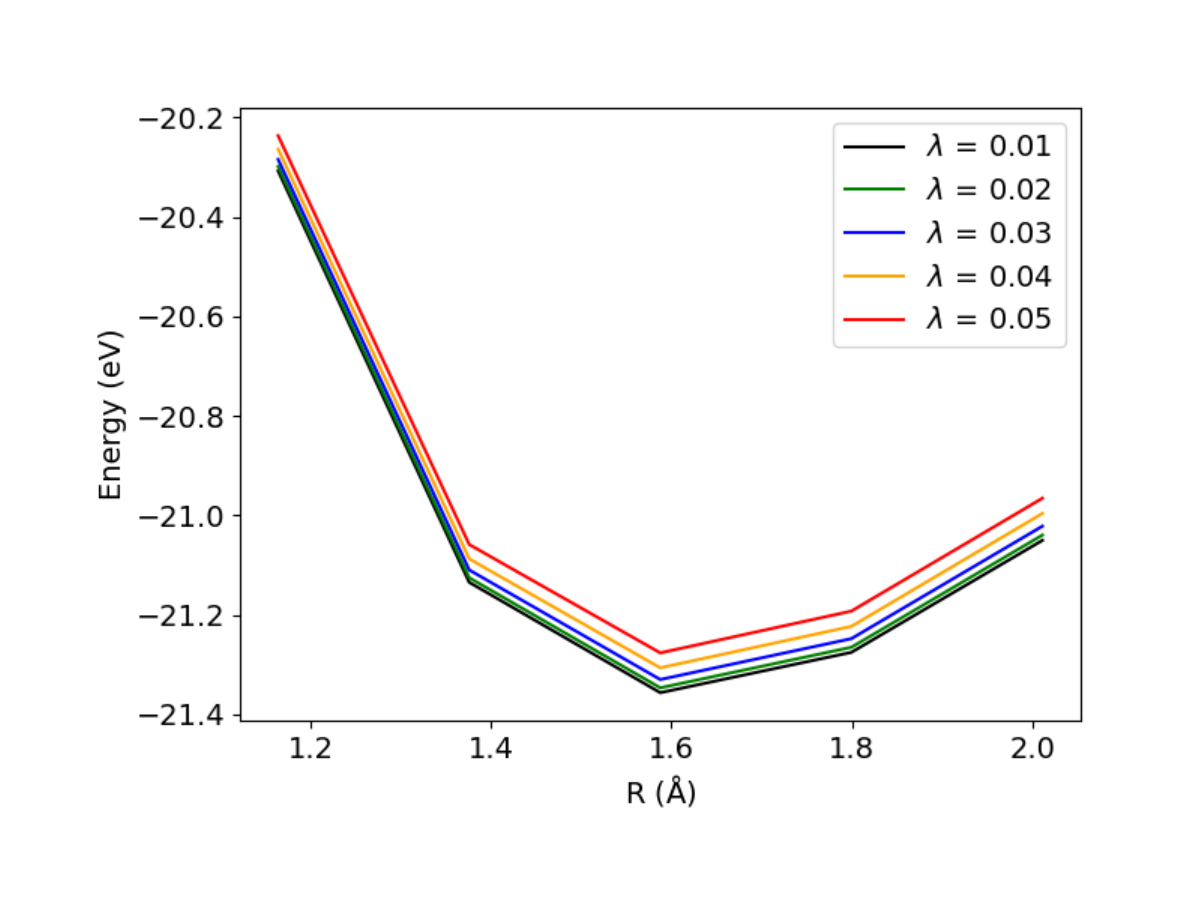}
\caption{LiH: Ground-state energy vs bond length for $\omega$
$=0.121$ for $\lambda=0.01,0.02,0.03,0.04,$ and $0.05$.}
 \end{subfigure}
 \hfill
 \begin{subfigure}[t]{0.48\textwidth}
\includegraphics[width=\linewidth]{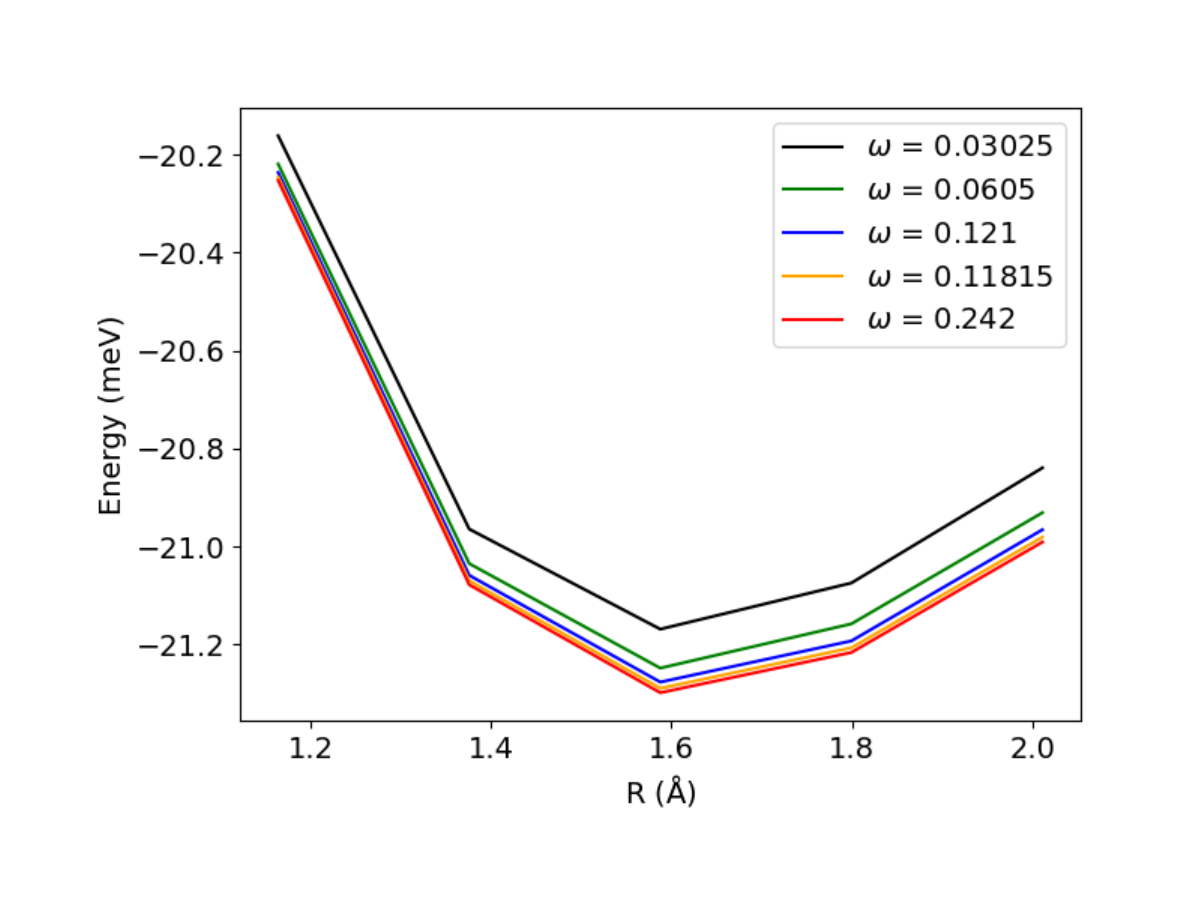}
\caption{LiH: Ground-state energy vs bond length for
$\omega=0.03025,0.0605,0.121,0.1815,$ and $0.242$ while keeping $\lambda=0.05$.
The grid parameters used in the calculation
are {$N_{x}=N_{y}=N_{z}=101$} and $h=0.2$ a.u.\ grid spacing in both
cases.}
 \end{subfigure}
 \caption{}
 \label{fig:lih1}
\end{figure*}

\begin{figure*}[t]
 \centering
 \begin{subfigure}[t]{0.48\textwidth}
\includegraphics[width=\linewidth]{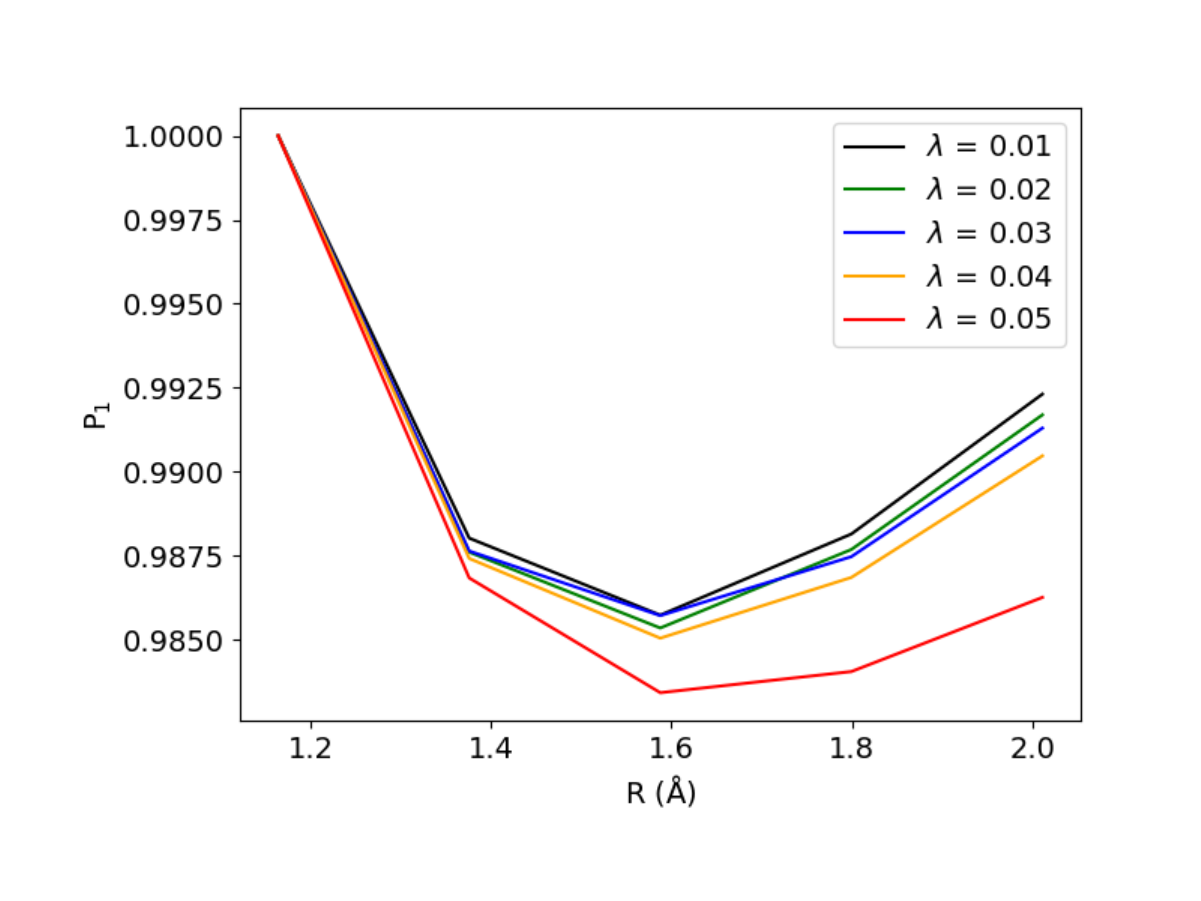}
\caption{LiH: Occupation of the {$\ket{1}$} space vs bond length for $\omega$
$=0.121$ for $\lambda=0.01,0.02,0.03,0.04,$ and $0.05$. The occupation
values are divided by 0.00073, 0.00289, 0.00644, 0.01126{,} and
0.01722 to fit the curves in the same figure. In other words, the
occupation for $\lambda=0.01$ is of the order of 0.00073, and that of
$\lambda=0.05$ is of 0.0172, respectively. }
 \end{subfigure}
 \hfill
 \begin{subfigure}[t]{0.48\textwidth}
\includegraphics[width=\linewidth]{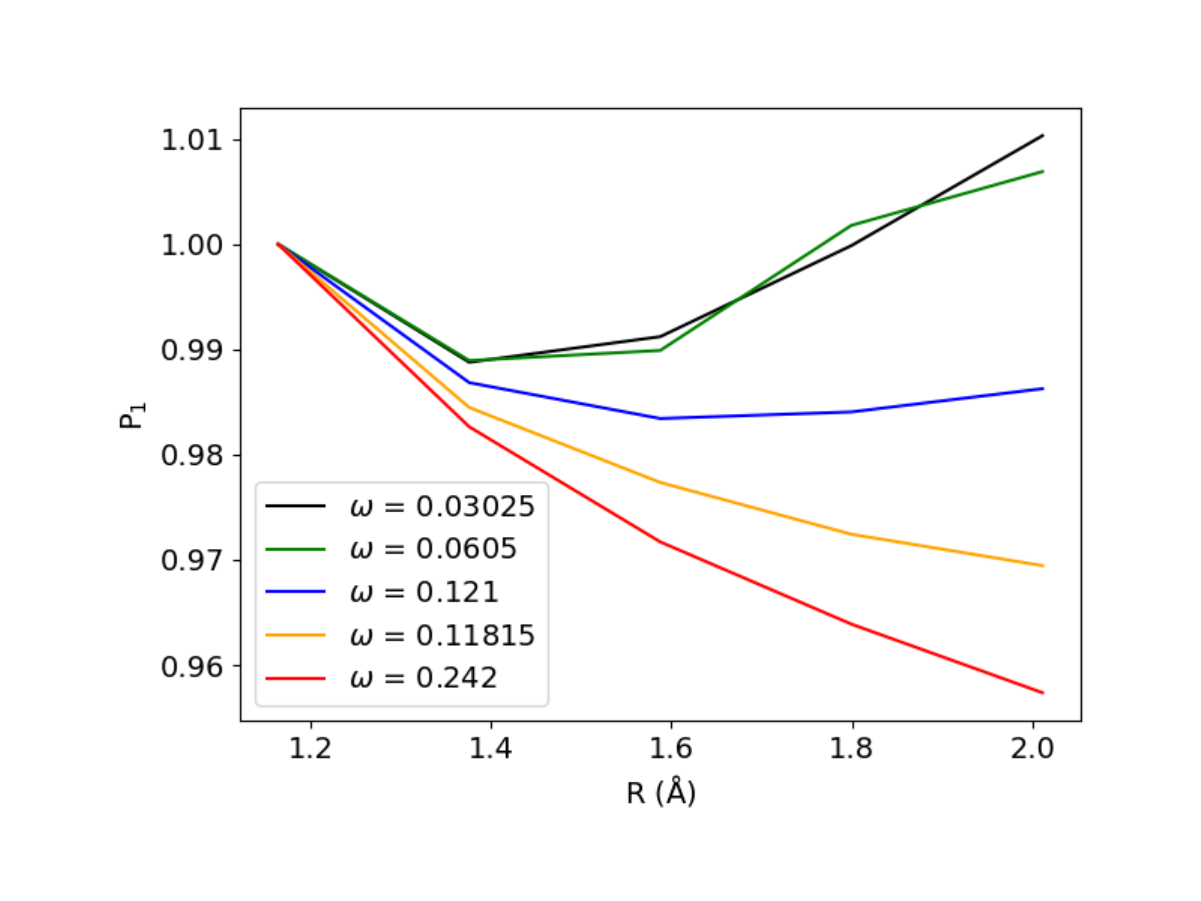}
\caption{LiH: Occupation of the {$\ket{1}$} space vs bond length for
$\omega=0.03025,0.0605,0.121,0.1815,$ and $0.242$ while keeping
$\lambda=0.05$. The occupation values are divided by
0.0961, 0.0460, 0.0172, 0.0087{, and} 0.0051 to fit the curves on the same
figure.}
 \end{subfigure}
 \caption{}
 \label{fig:lih2}
\end{figure*}

Fig.~\ref{fig:lih1} presents the ground-state energy of LiH as a
function of bond length for various coupling strengths $\lambda
$ and cavity frequencies $\omega
$. The molecular system is coupled to both the $\ket{0}$
and $\ket{1}$ photon Fock states.

The results demonstrate that increasing $\lambda
$ elevates the ground-state energy due to the diamagnetic contribution
in the Pauli--Fierz Hamiltonian. Notably, the overall shape of the
potential energy curves remains qualitatively unchanged across
different coupling strengths, exhibiting primarily a vertical shift
that scales approximately as 2$\lambda^2
$. This behavior indicates that the diamagnetic self-energy
contributes a nearly constant offset across all bond lengths.

In contrast, increasing $\omega$ reduces the ground-state energy (Fig.~\ref{fig:lih1}), 
producing an opposite effect on the system's energetics. This
inverse relationship arises because the diamagnetic term, which
dominates the energy modification, exhibits an inverse dependence on
the cavity frequency.
Fig.~\ref{fig:lih2} illustrates the variation in $\ket{1}$
state occupation, which quantifies the photonic excitation from
the $\ket{0}$ state. The bond-length dependence of the $\ket{1}$
occupation follows a trend similar to that observed for the energy
in Fig.
\ref{fig:lih1}. The occupation increases substantially with larger
$\lambda$ values; note that the occupations in Fig.
\ref{fig:lih2} are scaled by different multiplicative factors to
facilitate visual comparison on a single plot.
The cavity frequency dependence of the $\ket{1}$
occupation (Fig.~\ref{fig:lih2}) exhibits behavior similar to the energy variation for
$\omega = 0.121$~a.u. and $\omega = 0.1815
$~a.u. However, for larger $\omega$
values, the occupation decreases with increasing internuclear
separation.

\subsection{BH$_3$}

In this section, we compare our calculated Rabi splitting for BH$_3$
 with QED-FCI results from Ref.~\cite{vu2024cavity}. While the
QED-FCI approach determines the Rabi splitting by directly
diagonalizing the Hamiltonian to obtain the energy separation between
upper and lower polariton states, our methodology employs an
alternative strategy. We compute the absorption spectrum via real-time
propagation and extract the Rabi splitting from the spectral positions
of the polaritonic peaks.

Our TDDFT time-propagation approach evaluates the absorption spectrum
by applying a brief delta-function electric field pulse to the
ground-state system, then propagating the time-dependent Kohn--Sham
equations to track the subsequent evolution of the electronic density.
The time-dependent induced dipole moment is recorded throughout the
propagation, from which we obtain the frequency-dependent
polarizability through Fourier transformation. The absorption spectrum
is then derived from the imaginary component of the polarizability,
which corresponds directly to the optical absorption cross-section.
The Rabi splitting is identified as the frequency separation between
the upper and lower polariton peaks in this spectrum.

\begin{figure*}[t]
\centering
\begin{subfigure}[t]{0.49\textwidth}
 \includegraphics[width=\textwidth]{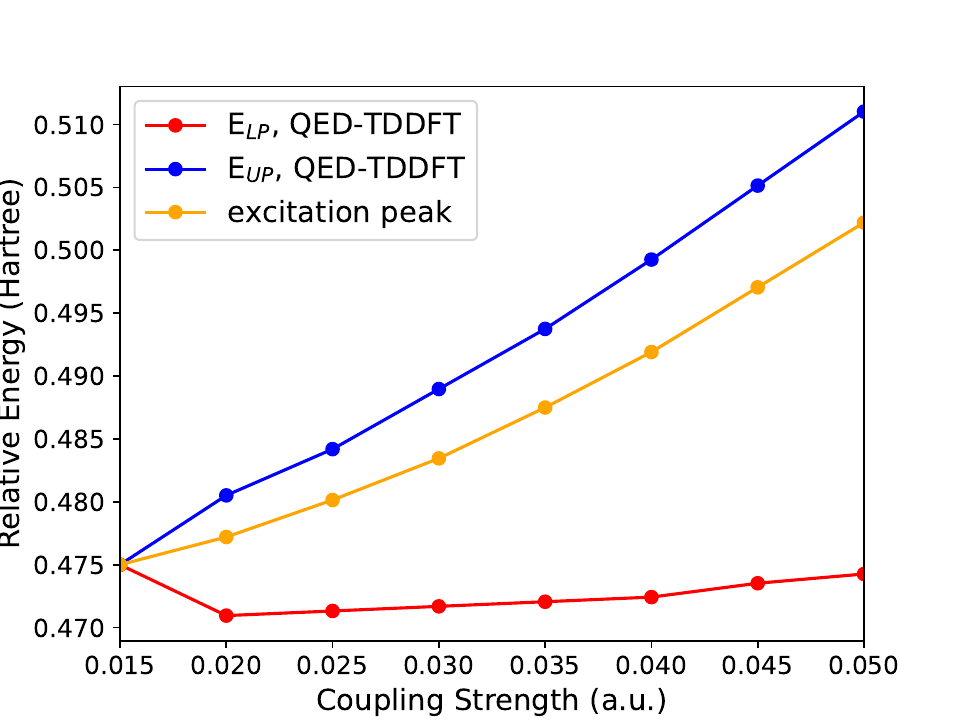}
\end{subfigure}
\hfill
\begin{subfigure}[t]{0.49\textwidth}
 \includegraphics[width=\textwidth]{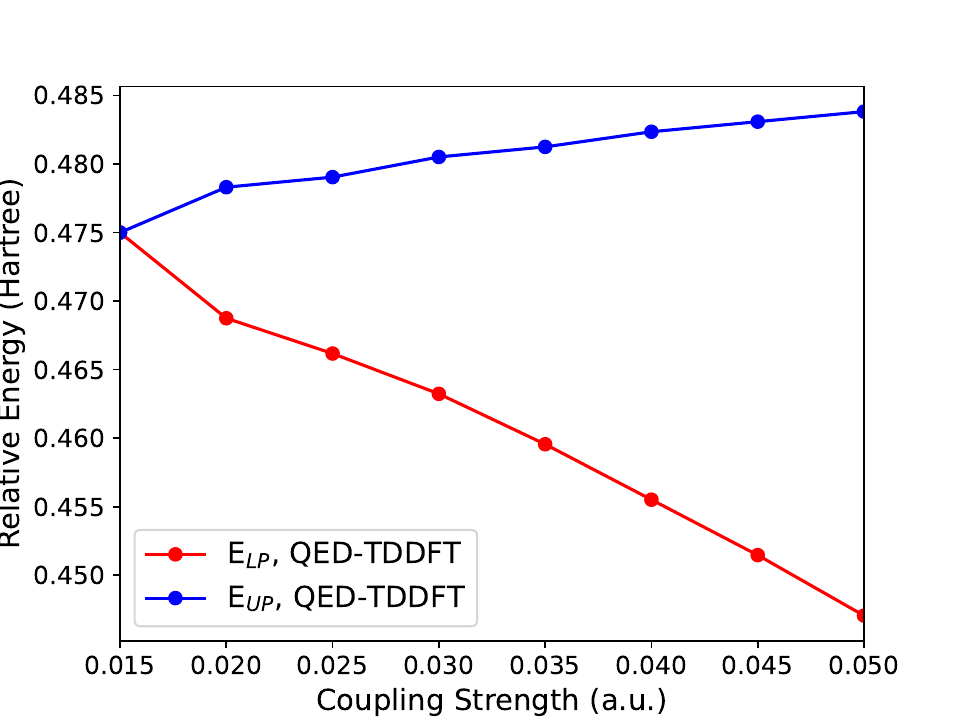}
\end{subfigure}
\caption{Left: Dependence of the polaritonic energies
on $\lambda$. The upper and lower {curves} show the energies of the
upper and lower polaritons{;} the middle curve {shows} the energy
dependence due to the diamagnetic term without coupling to the light.
Right: Upper and lower polaritonic energies {after subtracting} the
diamagnetic term's effect.
The parameters used for the calculation are {$N_{x}=N_{y}=N_{z}=71$}, {$h=0.3$ a.u.} grid spacing,
{$\Delta t=0.01$ a.u.} and {the} number of time steps {is} 100000.}
\label{fig:4}
\end{figure*}

\begin{figure}[t]
\centering
\includegraphics[width=0.9\linewidth]{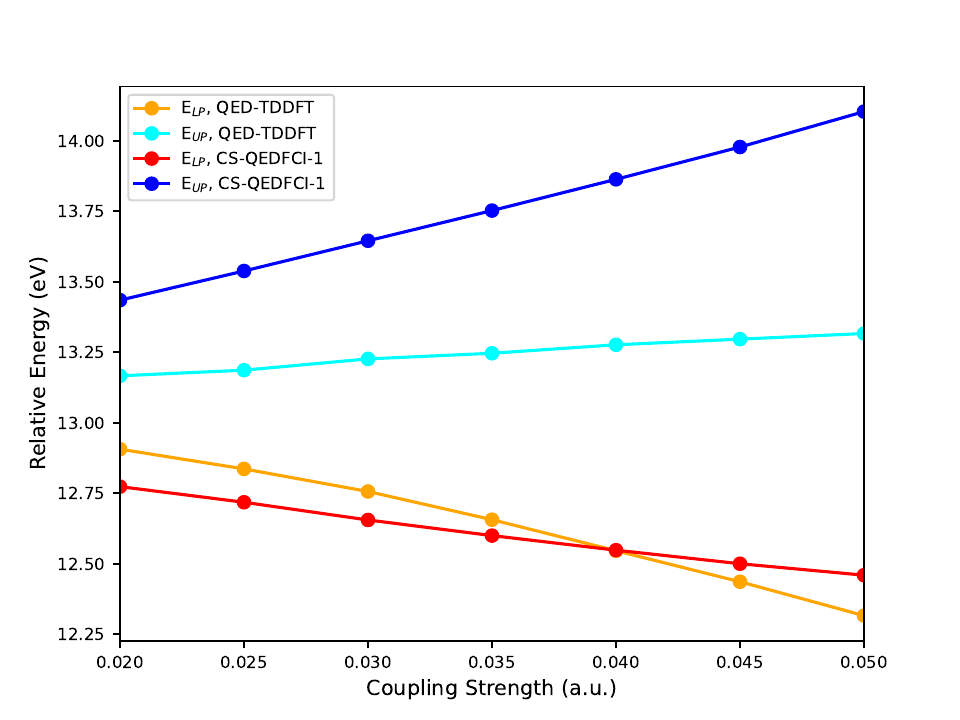}
\caption{A {comparison} of the polaritonic energies of BH{$_3$} using QED-TDDFT and QED-FCI.
The QED-TDDFT graph was shifted to fit it into the same scale. The graph begins at $\lambda=0.02$
because for lower coupling strengths we do not resolve a splitting in QED-TDDFT.
The same TDDFT parameters were used as in Fig.~{\ref{fig:4}}.}
\label{fig:5}
\end{figure}
The cavity-free case exhibits an excitation peak at {$0.4732$ a.u.} (see Fig.
\ref{fig:4}), which closely matches the molecule's third singlet excited state
reported by Vu et al.~(2024) \cite{vu2024cavity}. Initially, we couple the molecule to the
{$\ket{0}$} Fock-state. Although this configuration lacks light{--}matter
coupling, the excitation peak energy increases due to the positive
diamagnetic term, as demonstrated in Fig.~\ref{fig:4}.
To determine the upper and lower peak positions, we set the cavity
frequency $\omega$ equal to $E/\hbar$, where $E$ represents the absorption peak
energy (middle curve in Fig.~\ref{fig:4}). Since the diamagnetic term causes the peak
position to shift with varying $\lambda$, we correspondingly adjust
$\omega$ in our
calculations. When light{--}matter coupling is introduced, the single
absorption peak splits into two distinct peaks. The energies of these
split peaks define the upper and lower polaritons.

Fig.~\ref{fig:4} illustrates the $\lambda$ dependence beginning at
$\lambda = 0.015$. Below
this threshold value, peak splitting is not observed. This limitation
arises because QED-TDDFT-TP produces absorption spectra with peaks of
finite width, causing overlap at small $\lambda$ values. While extending the
propagation time could potentially resolve this issue, such
calculations would be computationally impractical.
As anticipated, the upper polariton energy increases with $\lambda$. However,
the lower polariton energy exhibits minimal variation. This behavior
occurs because the peak frequencies shift upward with increasing
$\lambda$, creating the appearance that $E_{LP}$ remains constant. To provide a
clearer visualization, the right panel of Fig.~\ref{fig:4} shows the polariton
energies after removing the diamagnetic contribution (by subtracting the
middle curve from both the upper and lower curves). With this
adjustment, both upper and lower polaritonic energies display the
expected $\lambda$-dependent behavior.

Fig.~\ref{fig:5} presents a comparison of polaritonic energies for BH{$_3$}
calculated using QED-TDDFT and QED-FCI methods for the same excited
state. The QED-FCI results are taken from Vu et al.~(2024)
\cite{vu2024cavity}. The graph reveals that while both methods demonstrate increasing
relative energy with rising $\lambda$ values, QED-TDDFT-TP consistently
underestimates the splitting magnitude and exhibits greater nonlinear
behavior compared to QED-FCI. Additionally, our approach
underestimates the upper polariton energy. {Given the sensitivity of QED-FCI to active-space size noted in Ref.~\cite{vu2024cavity}, the overall agreement is encouraging.}

\subsection{(H{$_2$}){$_2$}}
\begin{figure}[t]
\centering
\includegraphics[width=\linewidth]{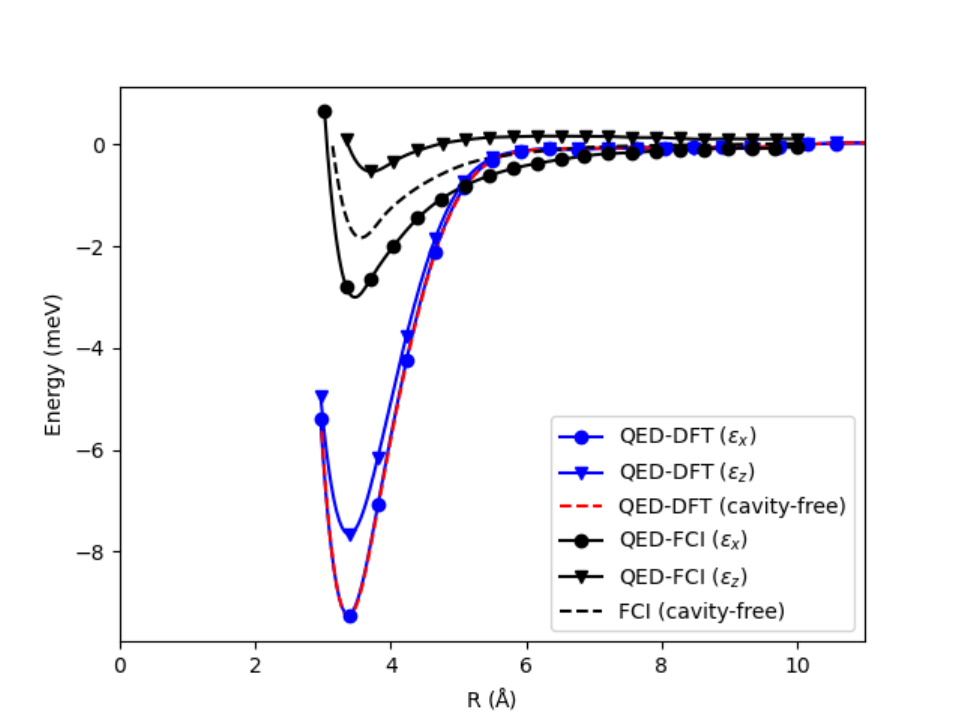}
\caption{Potential energy surface for (H{$_2$}){$_2$}. $\lambda = 0.05$, $\omega = 12.7$ eV.}
\label{fig:6}
\end{figure}

Intermolecular forces can be fundamentally understood as electronic
interactions mediated by transverse electromagnetic fields
\cite{craig1998molecular,haugland2021intermolecular}. Consequently,
modifying electromagnetic boundary conditions through cavity
confinement can substantially alter intermolecular interactions
\cite{Salam2016NonRelativistic}. Moreover, in the strong coupling
regime, molecules can interact with one another via delocalized cavity
photons even at separations where direct Coulombic interactions become
negligibly weak.
Fig.~\ref{fig:6} presents the potential energy surface of (H$_2$)$_2$ 
computed using QED-DFT-TP, compared with QED-FCI results from
Ref.~\cite{haugland2021intermolecular}. The QED-DFT-TP method
reproduces the overall topology of the QED-FCI surface and correctly
predicts that cavity modes with $\epsilon_z$
 polarization (parallel to the intermolecular axis) yield lower
binding energies than those with $\epsilon_x$
polarization (perpendicular). However, QED-DFT-TP systematically
overestimates the binding energy across all configurations.

A notable discrepancy is that QED-DFT-TP predicts the $\epsilon_x$
polarization to produce a potential energy surface nearly identical
to the cavity-free case, whereas QED-FCI shows substantial cavity
modification for this polarization. It is important to recognize that
in the cavity-free limit, our method reduces to conventional DFT,
which cannot accurately describe van der Waals interactions despite
yielding qualitatively reasonable curves
\cite{klimevs2012perspective}. We also note that the QED-DFT approach
employed in Ref.~\cite{haugland2021intermolecular} does not appear to
reproduce a bound potential energy minimum for this system.

\subsection{Ar{$_2$}}
\begin{figure}[t]
\centering
\begin{subfigure}[t]{0.49\textwidth}
 \includegraphics[width=\textwidth]{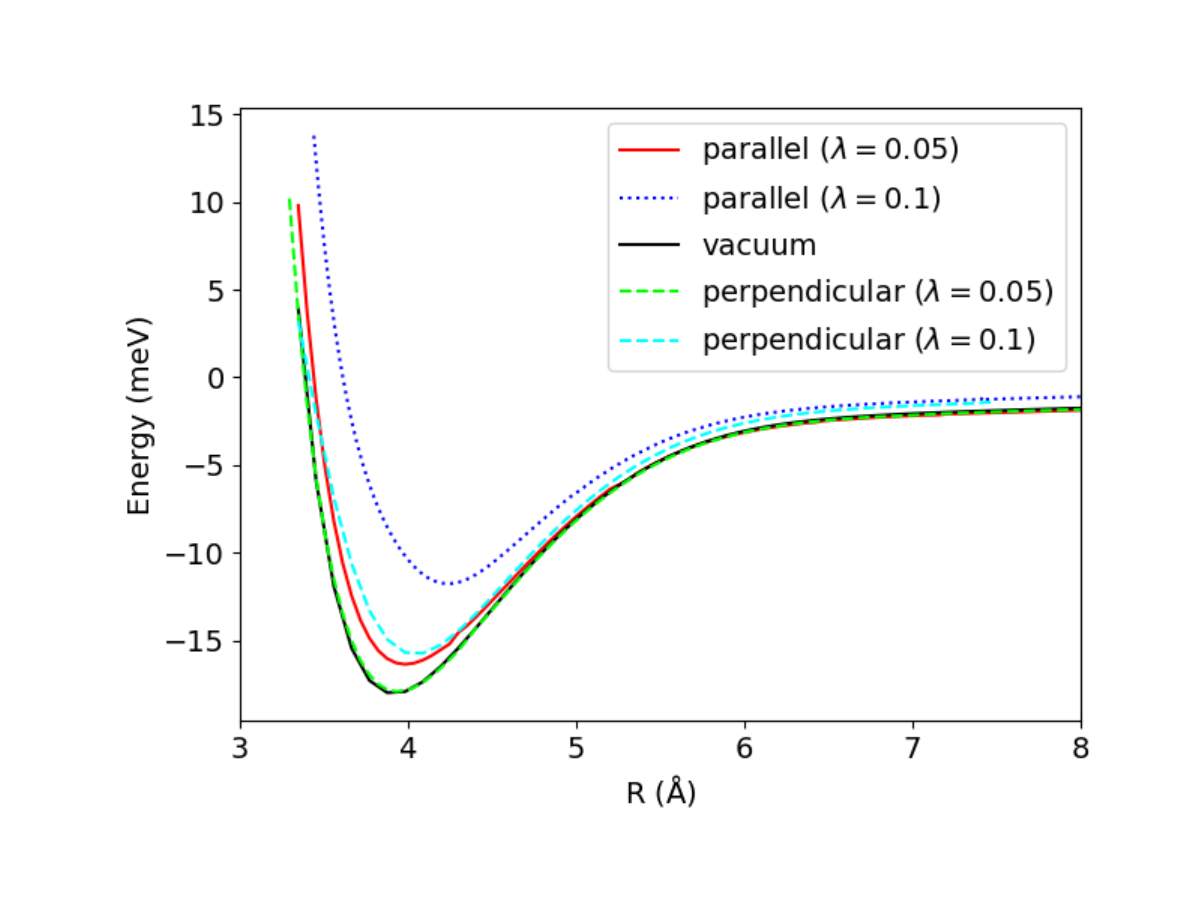}
\end{subfigure}
\begin{subfigure}[t]{0.49\textwidth}
 \includegraphics[width=\textwidth]{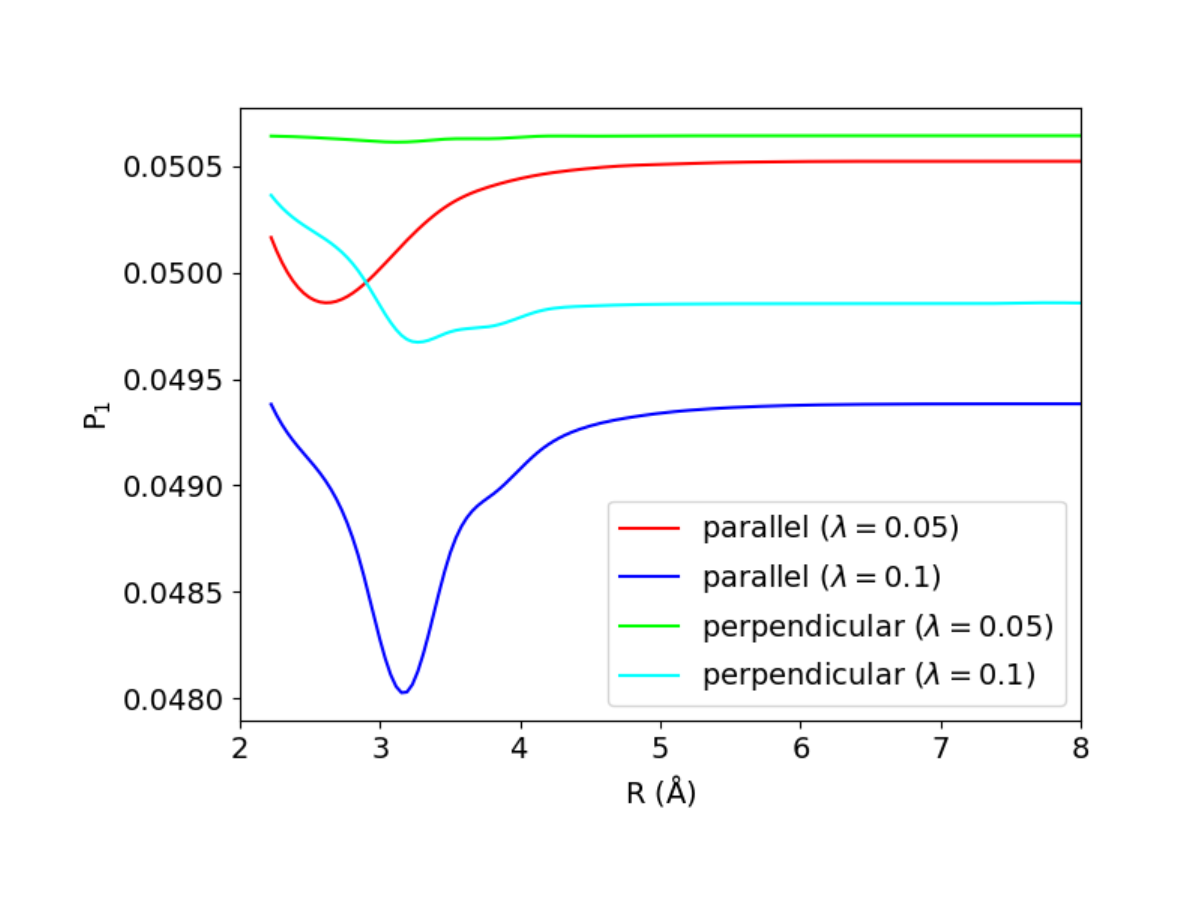}
\end{subfigure}
\caption{Potential energy surface (top) and occupation probability
of the {$\ket{1}$} Fock-space (bottom) for Ar{$_2$}.
To enable comparison within a single figure, the occupation
probabilities corresponding to $\lambda=0.05$ have been offset by adding
0.38. {$\omega=0.467$ a.u.\ was} used in the calculations.}
\label{fig:71}
\end{figure}

We have also investigated cavity-modified intermolecular interactions
in Ar$_2$. The two argon atoms are positioned along the 
$z$-axis with varying separation distance $R$. In the cavity-free case ($\lambda=0
$), our results closely match those of Ref.
\cite{PhysRevLett.134.073002}, predicting an equilibrium separation of
3.9\AA, although our binding energy is approximately 5~meV stronger.
As shown in Fig.\ref{fig:71}, the binding energy remains nearly
unchanged when the cavity polarization $\boldsymbol{\epsilon}$
 is perpendicular to the dimer axis and the coupling strength is weak
($\lambda=0.05$). This behavior can be understood by examining the occupation
probability of the $\ket{1}$
Fock state (Fig.
\ref{fig:71}), which exhibits minimal variation with internuclear
separation, leaving the intermolecular potential essentially
unmodified relative to the cavity-free case.
Increasing the coupling strength to $\lambda=0.1$
introduces a stronger distance dependence in the $\ket{1}$
state occupation (Fig.~\ref{fig:71}), resulting in reduced binding energy compared to the cavity-free
reference.When the cavity polarization is parallel to the dimer axis, the
occupation of the $\ket{1}$ state exhibits pronounced sensitivity to the interatomic
separation, and the binding energy decreases substantially with
increasing $\lambda$.

\begin{figure}[t]
 \includegraphics[width=0.5\textwidth]{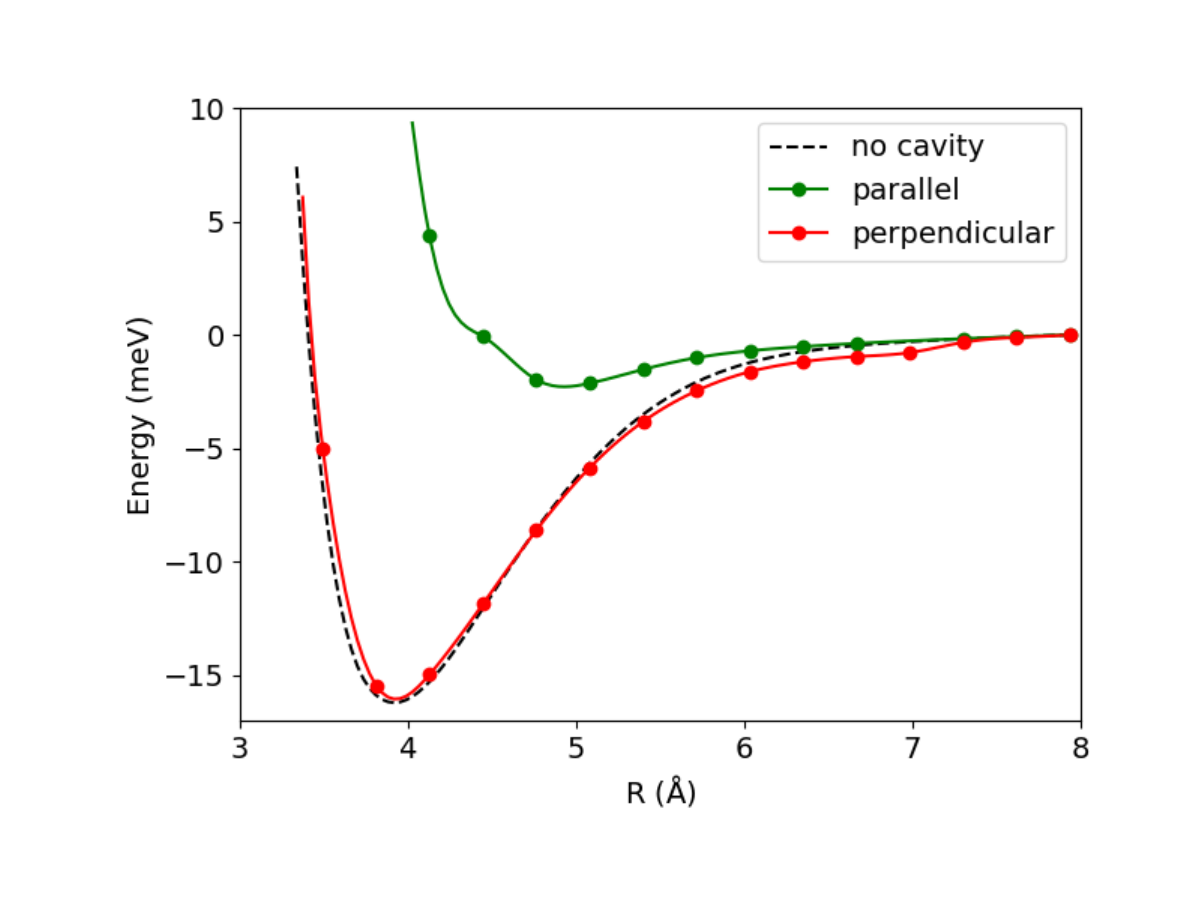}
\caption{Potential energy surface for Ar{$_2$}. $\lambda=0.1$ and
$\omega=0.0375$ a.u.\ used in the calculations.}
\label{fig:72}
\end{figure}

Fig.~\ref{fig:72} displays the Ar dimer energy as a function of interatomic
distance using parameters $\lambda=0.1$
and $\omega=0.0375$ atomic units, identical to those employed in Ref.~\cite{PhysRevLett.134.073002}.
When compared to Fig.~\ref{fig:71}, the perpendicular configuration exhibits
enhanced binding energy that approaches the cavity-free binding energy
more closely, which results from the tenfold reduction in frequency.
Conversely, the parallel configuration shows reduced binding relative
to the previous case. Ref.~\cite{PhysRevLett.134.073002} predicted that parallel
configurations would be less bound than the cavity-free case, while
perpendicular configurations would be more strongly bound than the
cavity-free case. Our findings are consistent with Ref.~\cite{PhysRevLett.134.073002}
regarding the parallel configuration but disagree concerning the
perpendicular configuration.

\subsection{(H{$_2$}O){$_2$}}
\begin{figure}[t]
\centering
\includegraphics[width=\linewidth]{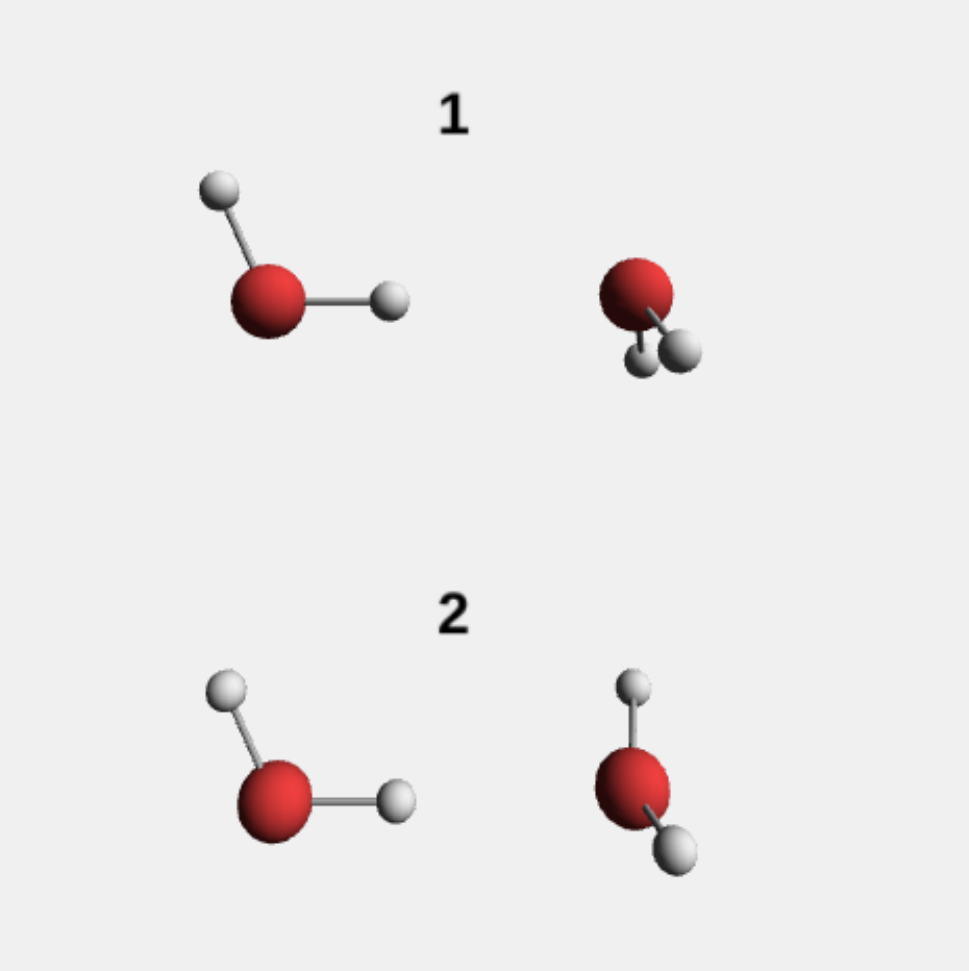}
\caption{The two water dimer configurations used in the calculations.}
\label{fig:81}
\end{figure}

We investigated the influence of cavity confinement on hydrogen
bonding in a water dimer by varying the oxygen--oxygen separation
distance RR
R. This system was recently examined in Ref.
\cite{haugland2021intermolecular}. Since the exact atomic coordinates
from that study were not available, we constructed two geometrically
similar configurations based on the structures depicted in
Ref.\cite{haugland2021intermolecular}, as shown in Fig.\ref{fig:81}.
The cavity polarization is oriented along the O--O axis, and the
cavity frequency and coupling strength are set to match those of
Ref.\cite{haugland2021intermolecular}: $\omega = 7.86
$~eV and $\lambda = 0.1$.

\begin{figure}[t]
\centering
\includegraphics[width=\linewidth]{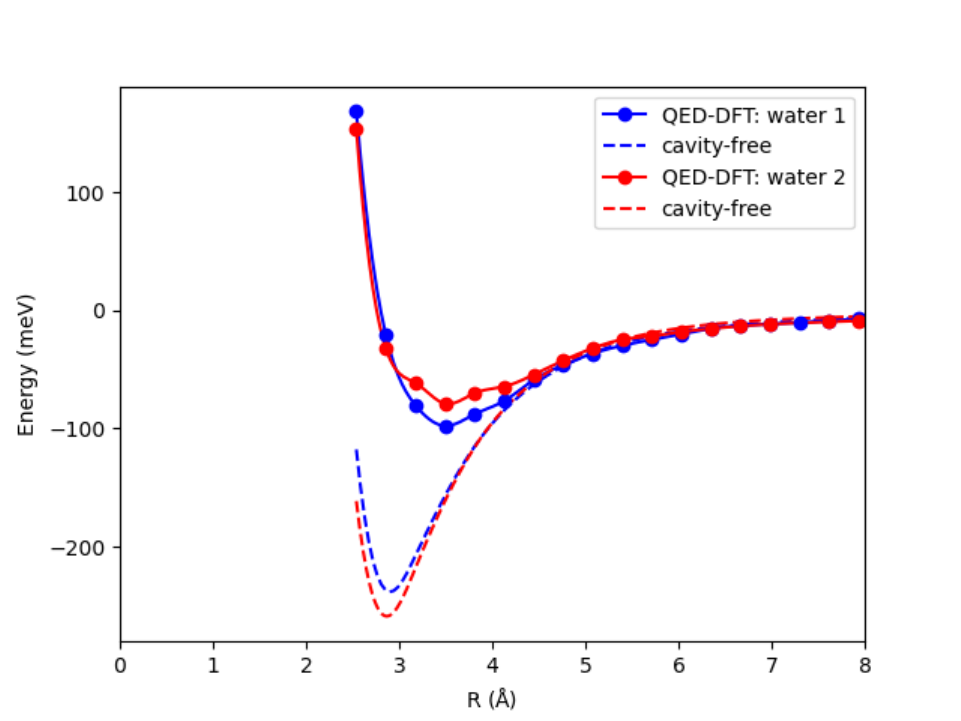}
\caption{Potential energy surfaces for (H{$_2$}O){$_2$} systems.}
\label{fig:8}
\end{figure}

Fig.~\ref{fig:8} shows that our potential energy curve exhibits
excellent agreement with the Coupled-Cluster Singles and Doubles
(CCSD) results from Ref.\cite{haugland2021intermolecular} in the
cavity-free case. Both CCSD and QED-DFT-TP predict an equilibrium
separation of 2.9\AA. The binding energies are also in good agreement:
CCSD yields 214 meV (Fig.6 in Ref.\cite{haugland2021intermolecular}),
while QED-DFT-TP predicts 240 meV.
Under cavity confinement, CCSD calculations indicate a modest 30meV
reduction in binding energy, whereas QED-DFT-TP predicts a
substantially larger decrease of approximately 130 meV. This more
pronounced modification is physically reasonable given the strong
light--matter coupling regime accessed by the chosen parameters. The
quantitative discrepancy between the two approaches likely stems from
differences in the water dimer geometries employed. This
interpretation is supported by the substantial variation in
cavity-induced binding energy changes observed between our two
distinct water dimer configurations, which underscores the geometric
sensitivity of cavity effects on hydrogen bonding.

\subsection{HF dimer}
\begin{figure}[t]
\centering
\includegraphics[width=\linewidth]{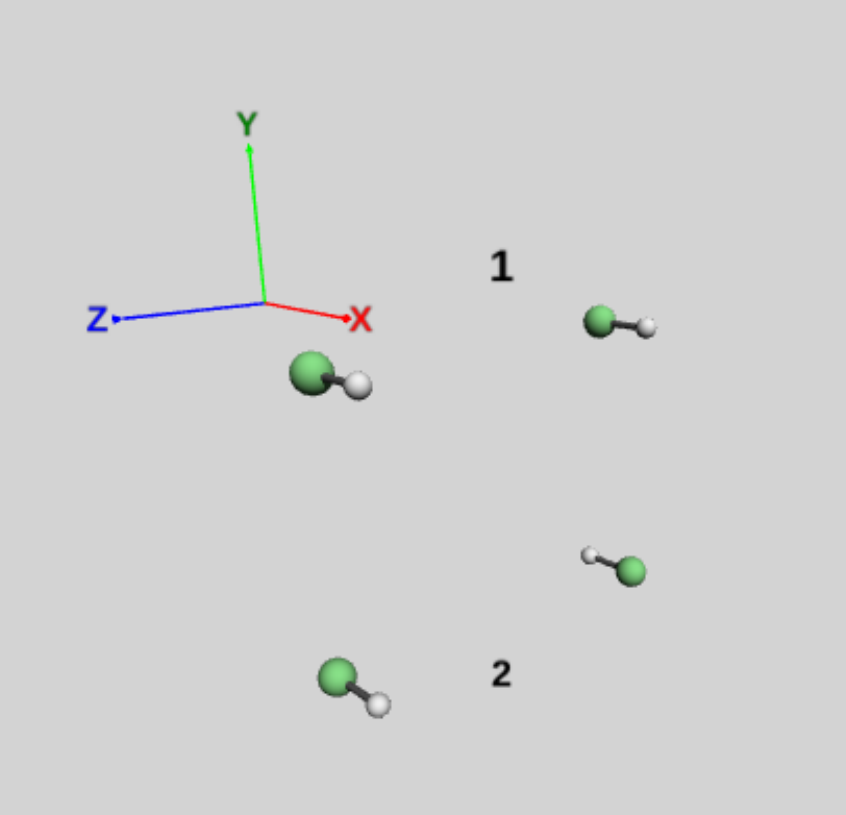}
\caption{The two HF dimer configurations used in the calculations.}
\label{fig:9}
\end{figure}

We examined the distance-dependent binding energy of HF dimers in two
distinct molecular arrangements: parallel orientation (HF--HF) and
antiparallel orientation (HF--FH), as illustrated in Fig.~\ref{fig:9}.
The HF molecular axes are oriented along the $x$-direction. 
Two cavity polarization configurations were investigated:
$\boldsymbol{\epsilon} = (1,0,0)$, designated as ``perpendicular'' 
since it is orthogonal to
the intermolecular $z$-axis, and $\boldsymbol{\epsilon} = (0,0,1)$, 
termed ``parallel'' as it aligns with the axis connecting
the two molecules.

\begin{figure}[t]
\centering
\begin{subfigure}[t]{0.49\textwidth}
 \includegraphics[width=\textwidth]{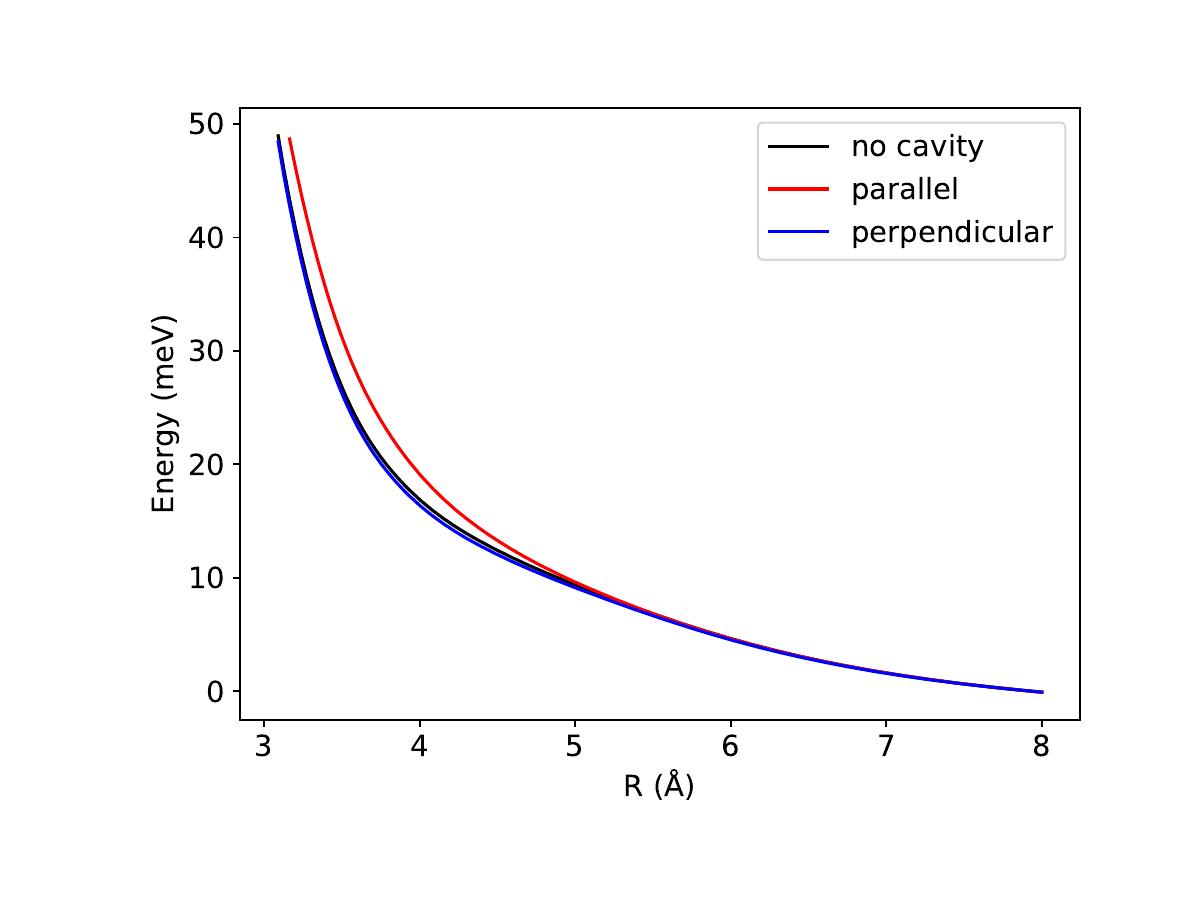}
\end{subfigure}
\begin{subfigure}[t]{0.49\textwidth}
 \includegraphics[width=\textwidth]{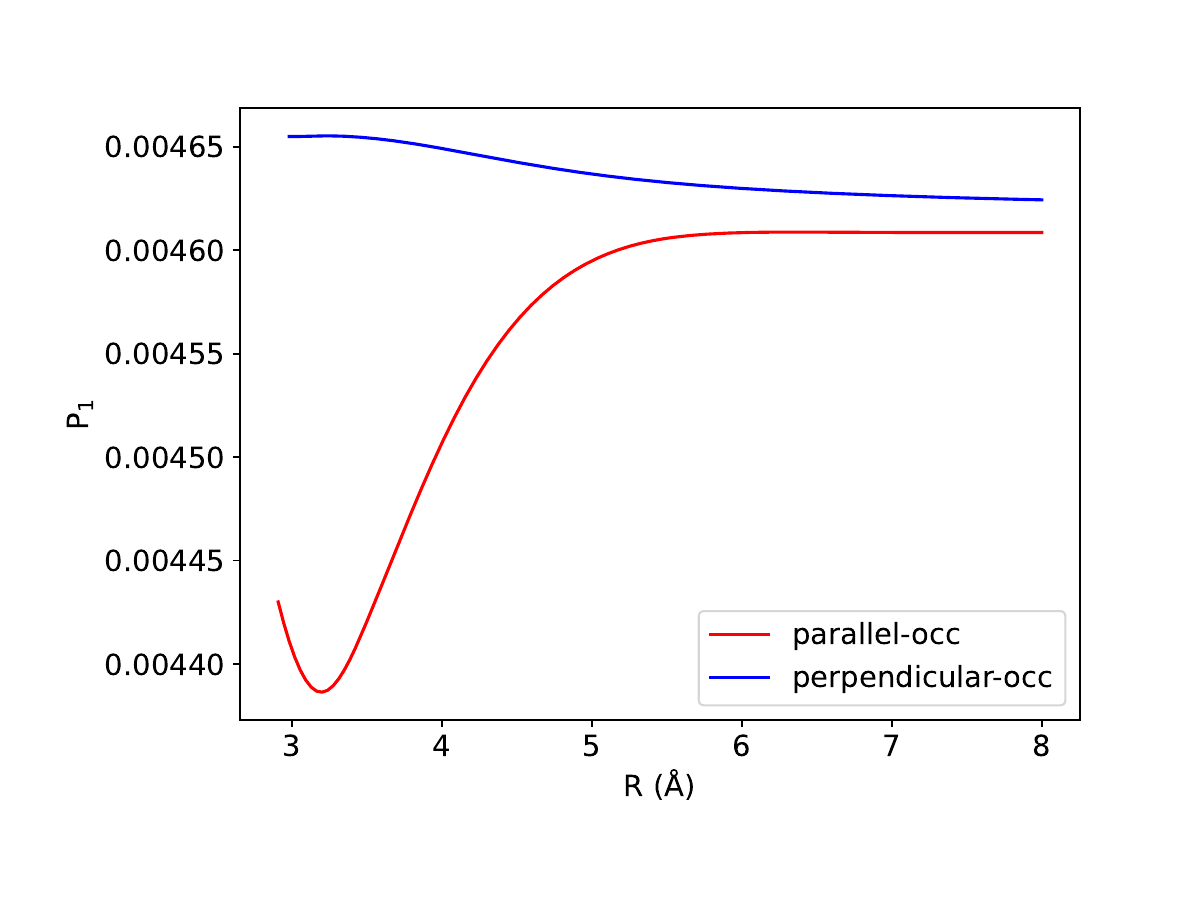}
\end{subfigure}
\caption{Potential energy surface (top) and occupation probability
of the {$\ket{1}$} Fock-space (bottom) for the HF dimer
(dimer 1 in Fig.~\ref{fig:9}).}
\label{fig:10}
\end{figure}

For the parallel HF--HF configuration, Fig.~\ref{fig:10} demonstrates
that the cavity-free system exhibits no intermolecular binding, and
light--matter coupling in either polarization fails to induce
molecular association. The parallel polarization yields higher total
energies than the perpendicular case, while both polarizations produce
minimal $\ket{1}$ photon-state occupation across all intermolecular separations.

\begin{figure}[t]
\centering
\begin{subfigure}[t]{0.49\textwidth}
 \includegraphics[width=\textwidth]{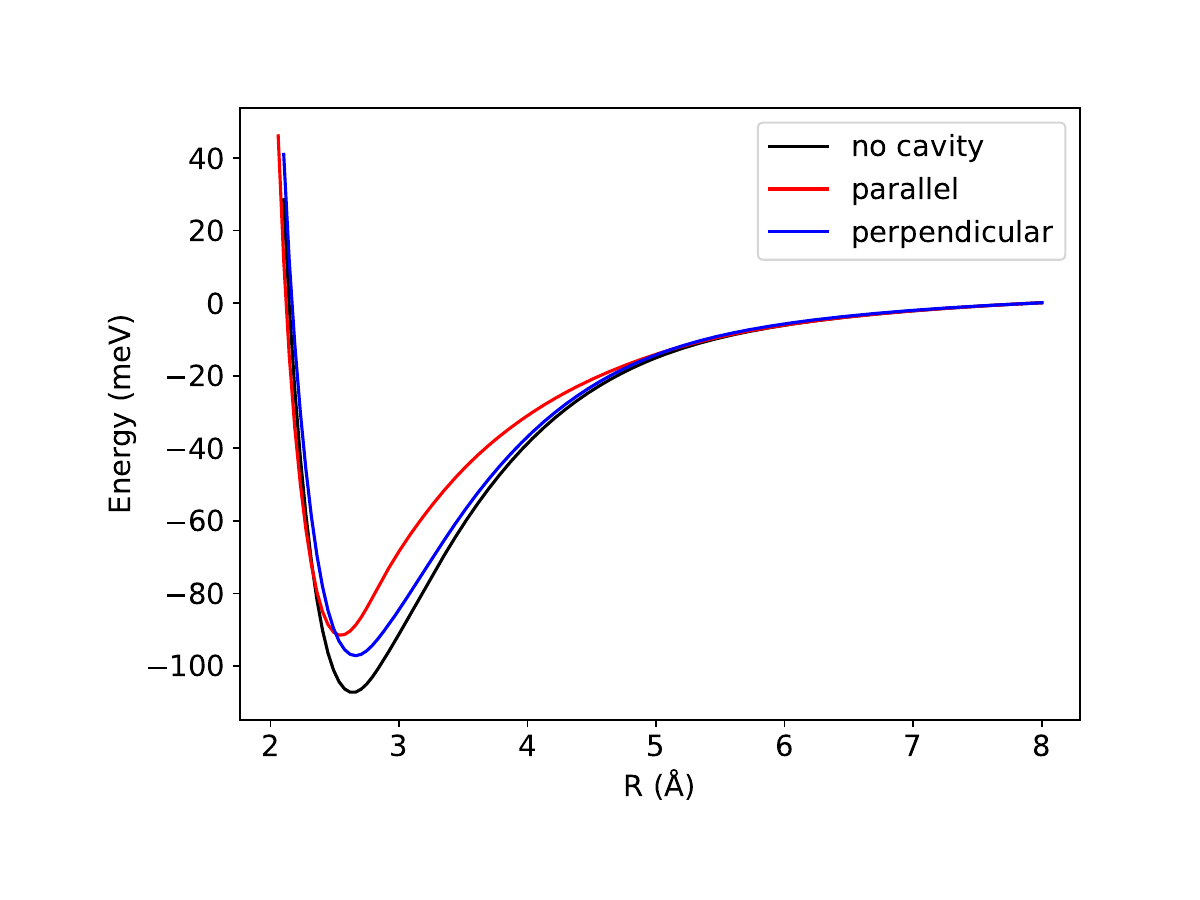}
\end{subfigure}
\begin{subfigure}[t]{0.49\textwidth}
 \includegraphics[width=\textwidth]{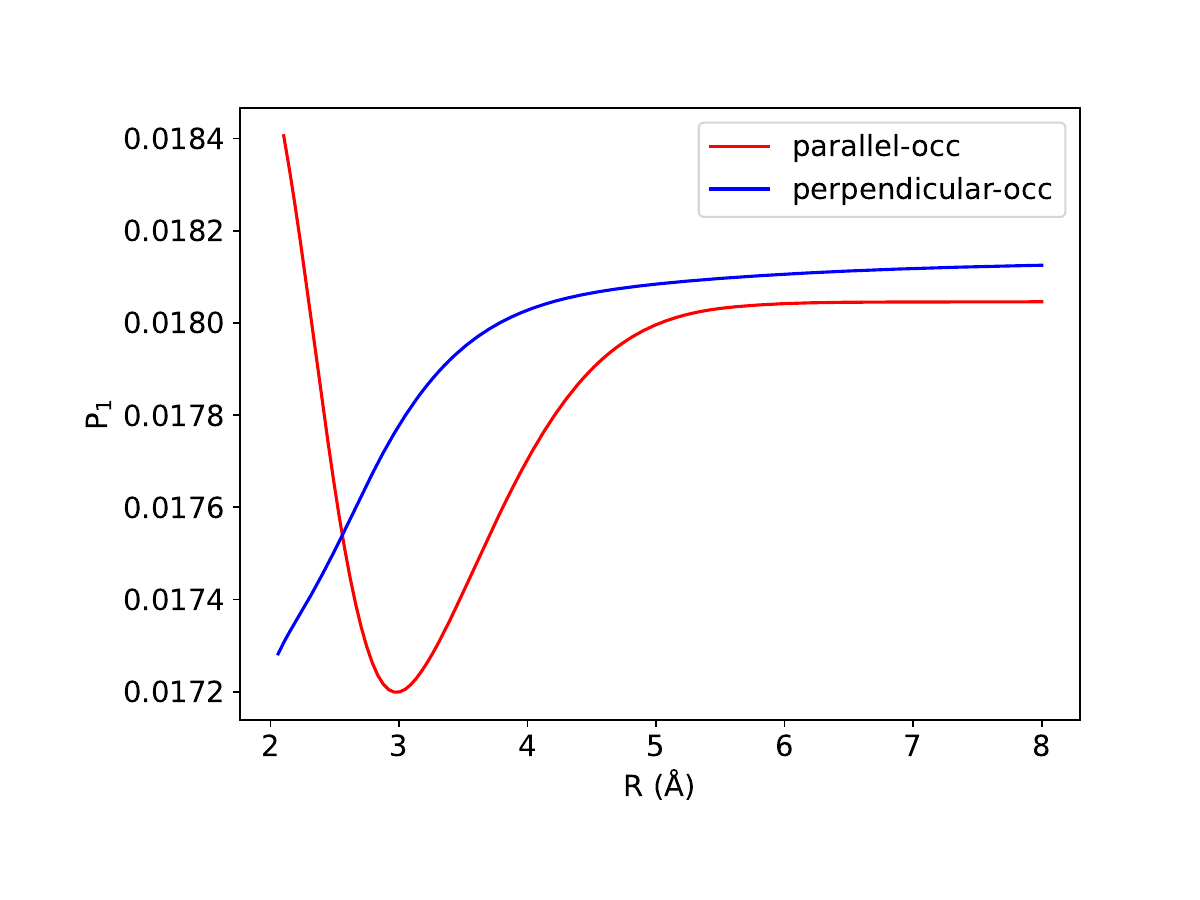}
\end{subfigure}
\caption{Potential energy surface (top) and occupation probability
of the {$\ket{1}$} Fock-space (bottom) for the HF dimer
(dimer 2 in Fig.~\ref{fig:9}).}
\label{fig:11}
\end{figure}

In contrast, the antiparallel HF--FH arrangement (Fig.\ref{fig:11})
exhibits substantial intermolecular binding with an energy minimum at
approximately 2.7\AA—notably about three times the HF monomer bond
length of 0.92~\AA. Light--matter coupling reduces the binding
strength for both polarization directions relative to the cavity-free
case. Parallel polarization decreases the equilibrium separation,
while perpendicular polarization increases it.
The $\ket{1}$ photon-state occupation in the antiparallel configuration is
significantly enhanced compared to the parallel arrangement. The
occupation profiles exhibit polarization-dependent behaviors: at short
intermolecular distances they differ substantially but converge at
large separations. Notably, the parallel polarization exhibits an
occupation minimum near 3~\AA, whereas the perpendicular polarization
shows monotonic behavior with no such feature across the investigated
distance range.


\section{Summary}
This work examines the effectiveness of combining quantum
electrodynamics with time-dependent density functional theory to model
molecular systems under strong coupling conditions within optical
cavities, benchmarking the results against existing theoretical
approaches. Conventional density functional theory is extended by incorporating
the Pauli{--}Fierz nonrelativistic quantum electrodynamics Hamiltonian,
with the coupled electron{--}photon system represented on a tensor
product of real-space and Fock-space.
The method is benchmarked against established wave function approaches
including photon-number quantum electrodynamical full configuration
interaction and complete active space configuration
interaction for small molecular systems. Potential
energy surfaces for LiH are shown to demonstrate excellent agreement
with reference calculations, while Rabi splitting calculations for BH{$_3$}
exhibit qualitative agreement despite some quantitative differences.
The influence of cavity parameters on intermolecular interactions is
investigated through studies of hydrogen-bonded and van der Waals
dimers, including (H{$_2$}){$_2$}, Ar{$_2$}, (H{$_2$}O){$_2$}, and (HF){$_2$}
systems. Cavity coupling
is found to significantly modify binding energies, equilibrium
distances, and photon occupation numbers, with strong dependence on
molecular orientation relative to field polarization observed. For HF
dimers, parallel (HF-HF) configurations are shown to remain unbound in
cavities while antiparallel (HF-FH) arrangements exhibit substantial
binding that decreases upon light{--}matter coupling.

The QED-TDDFT-TP approach demonstrates computational tractability
while maintaining sufficient accuracy to describe cavity QED effects
and polariton physics, reproducing the qualitative features of strong
coupling regimes. Residual quantitative discrepancies relative to
wave-function calculations warrant further analysis of
finite photon-basis effects and density-functional approximation quality.

The QED-DFT-TP method presented in the paper is orders of magnitude
cheaper than both QED-CC and especially QED-FCI. It retains explicit
quantized photon states while keeping computational cost close to
standard DFT, enabling simulations of realistic molecules and dimers
under strong light–matter coupling.

The present work assumes a bare matter exchange correlation
functional which is the same in each Fock-space. This approximation
could be a source of error relative to the exact solution. Future work
is needed to develop suitable exchange-correlation functionals for the
coupled electron orbitals Fock space approach.

\section{Acknowledgments}
This work was supported by the National
Science Foundation (NSF)
under Grant No. DMR-2217759.
Computational resources were provided by
ACES at
Texas A{\&}M University through allocation PHYS240167 from the
Advanced Cyberinfrastructure Coordination Ecosystem: Services {\&}
Support (ACCESS) program,
supported by NSF grants 2138259, 2138286, 2138307, 2137603, and 2138296.

\section*{Data Availability Statement}
The data that support the findings of this study are available
from the corresponding author upon reasonable request.

\section*{AUTHOR DECLARATIONS}
\par\noindent
{\bf Conflict of Interest}
The authors have no conflict of interest to disclose.

\appendix
\section{Length and {Velocity} gauge}
\label{appa}
The goal of this appendix is to give a brief overview of the different
forms of the Pauli{--}Fierz Hamiltonian and gauge invariance
\cite{rokaj2018light,DiStefano2019,PhysRevA.98.043801}.
Consider matter (atoms or molecules) interacting with quantized transverse radiation in the Coulomb gauge.
In the nonrelativistic limit and long-wavelength (dipole) approximation, the velocity-gauge Pauli{--}Fierz Hamiltonian reads
\begin{eqnarray}
\label{eq:Hv}
\hat H{_v}
&=& \sum{_{i=1}^{N_e}} \frac{1}{2{m_i}}\Bigl(\hat{\mathbf p}{_i} - {q_i}\,\hat{\mathbf A}\Bigr)^2
+ \hat V_{\mathrm{Coul+ext}}\nonumber \\
&+& \sum_{\alpha} \frac{\hbar\omega_\alpha}{2}\,\Bigl(\hat p_\alpha^2 + \hat q_\alpha^2\Bigr),
\end{eqnarray}
$\hat V_{\mathrm{Coul+ext}}$ collects Coulomb interactions and
external potentials{. Each} photonic mode $\alpha$ is a harmonic oscillator with canonical
pair $(\hat q_\alpha,\hat p_\alpha)$, obeying $[\hat q_\alpha,\hat p_\beta]=i\delta_{\alpha\beta}$.
In the long-wavelength limit, the vector potential is spatially uniform over the matter extent,
 \begin{equation}
 \hat{\mathbf A} = \sum_\alpha \mathcal A{_\alpha}\,\boldsymbol{\varepsilon}{_\alpha}\,\hat q{_\alpha},
 \qquad
 \mathcal A{_\alpha} \equiv \sqrt{\frac{\hbar}{{\varepsilon_0} V\,\omega_\alpha}},
 \end{equation}
with polarization $\boldsymbol{\varepsilon}{_\alpha}$, quantization volume $V$, and frequency $\omega{_\alpha}$.
Define the total matter dipole operator without charges,
\begin{equation}
\label{eq:R}
\hat{\mathbf R} \equiv \sum{_{i=1}^{N_e}} \mathbf r{_i} - \sum{_{j=1}^{N_n}} {Z_j} \mathbf R{_j},
\end{equation}
so that the physical dipole is $\hat{\mathbf D}=\sum{_i} {q_i}\, \mathbf r{_i} = -e\sum{_i} \mathbf r{_i} + \sum{_j} {Z_j} e\,\mathbf R{_j}$.
For brevity, many steps below are shown for a single electron with charge $-e$; the generalization is straightforward.
First we introduce the Power{--}Zienau{--}Woolley (PZW) transformation
\cite{doi:10.1098/rsta.1959.0008,Woolley1971}:
\begin{equation}
\label{eq:PZW}
\hat U \equiv \exp\!\Bigl[i\sum\alpha g{_\alpha}\,(\boldsymbol{\varepsilon}{_\alpha}\!\cdot\!\hat{\mathbf R})\,\hat q{_\alpha} \Bigr],
\qquad
g{_\alpha} \equiv \frac{e}{\sqrt{{\varepsilon_0} V\,\hbar\omega_\alpha}}.
\end{equation}
It effects simple shifts on the canonical operators:
\begin{equation}
\hat U^\dagger\,\hat q{_\alpha}\,\hat U = \hat q{_\alpha},
\end{equation}
\begin{equation}
\hat U^\dagger\,\hat p{_\alpha}\,\hat U = \hat p{_\alpha} + g{_\alpha}\,\boldsymbol{\varepsilon}{_\alpha}\!\cdot\!\hat{\mathbf R}, \label{eq:shift-phot}
\end{equation}
\begin{equation}
\hat U^\dagger\,\hat{\mathbf p}{_i}\,\hat U = \hat{\mathbf p}{_i} + e\,\hat{\mathbf A},
\end{equation}
\begin{equation}
\hat U^\dagger\,\hat{\mathbf A}\,\hat U = \hat{\mathbf A}, \label{eq:shift-matter}
\end{equation}
where we used $\hat{\mathbf A}=\sum\alpha \mathcal A{_\alpha} \boldsymbol{\varepsilon}{_\alpha} \hat q{_\alpha}$
and {$e\,\mathcal A_\alpha=\hbar g_\alpha$}
so that $e\hat{\mathbf A}=\hbar\sum\alpha g{_\alpha} \boldsymbol{\varepsilon}{_\alpha} \hat q{_\alpha}$.
Using \eqref{eq:shift-phot}{--}\eqref{eq:shift-matter} one finds
\begin{equation}
\hat U^\dagger\Bigl(\hat{\mathbf p}_i - (-e)\hat{\mathbf A}\Bigr)\hat U
= \hat{\mathbf p}_i,
\end{equation}
\begin{equation}
\hat U^\dagger \frac{\hbar\omega{_\alpha}}{2}\bigl(\hat p{_\alpha}^2 + \hat q_\alpha^2\bigr) \hat U
= \frac{\hbar\omega{_\alpha}}{2}\Bigl[\hat q{_\alpha}^2 + \bigl(\hat p{_\alpha} + g{_\alpha}\,\boldsymbol{\varepsilon}{_\alpha}\!\cdot\!\hat{\mathbf R}\bigr)^2\Bigr],
\end{equation}
while $\hat V_{\mathrm{Coul+ext}}$ remains invariant. Therefore,
\begin{eqnarray}
\label{eq:HL-qrep}
\hat H{_L}
&\equiv& \hat U^\dagger \hat H{_v} \hat U
= \sum{_{i=1}^{N_e}} \frac{\hat{\mathbf p}{_i}^2}{2{m_i}} + \hat
V_{\mathrm{Coul+ext}} \nonumber\\
&+& \sum\alpha \frac{\hbar\omega{_\alpha}}{2}\Bigl[\hat q{_\alpha}^2 + \bigl(\hat p{_\alpha} + g{_\alpha}\,\boldsymbol{\varepsilon}{_\alpha}\!\cdot\!\hat{\mathbf R}\bigr)^2\Bigr].
\end{eqnarray}
Equation \eqref{eq:HL-qrep} is the length-gauge Hamiltonian in the photonic coordinate representation.
Expanding the square makes the structure transparent:
\begin{eqnarray}
\label{eq:HL-decomposed}
\hat H{_L} &=& \sum{_i} \frac{\hat{\mathbf p}{_i}^2}{2{m_i}} + \hat
V_{\mathrm{Coul+ext}}
+ \sum\alpha \frac{\hbar\omega{_\alpha}}{2}\bigl(\hat q_\alpha^2 + \hat
p_\alpha^2\bigr)\nonumber \\
&\;-\;& \sum\alpha \hat{\mathbf D}\!\cdot\!\hat{\mathbf E}{_\alpha}
\;+\; \sum\alpha \frac{1}{2{\varepsilon_0}
V}\bigl(\boldsymbol{\varepsilon}_\alpha\!\cdot\!\hat{\mathbf D}\bigr)^2,
\end{eqnarray}
where the interaction is linear in the field (the $-\,\hat{\mathbf D}\!\cdot\!\hat{\mathbf E}$ coupling), and the last term is the dipole self-energy
(${e^2/(2\varepsilon_0 V)}$).
Here $\hat{\mathbf E}{_\alpha} \propto i\sqrt{\hbar\omega{_\alpha}/({\varepsilon_0} V)}\,\boldsymbol{\varepsilon}{_\alpha}(\hat a{_\alpha}-\hat a{_\alpha}^\dagger)$ is the transverse electric field of mode $\alpha$.

\subsubsection{Momentum representation and an equivalent form.}
A phase-space rotation of the photonic variables $(\hat q{_\alpha},\hat p{_\alpha})\mapsto (\hat p{_\alpha},-\hat q{_\alpha})$ (equivalently, $\hat a{_\alpha}\mapsto i\hat a{_\alpha}$) yields an equally valid representation:
\begin{eqnarray}
\label{eq:HL-prep}
\hat H{_L}
&=& \sum{_{i}} \frac{\hat{\mathbf p}{_i}^2}{2{m_i}} + \hat
V_{\mathrm{Coul+ext}} \nonumber\\
&+& \sum\alpha \frac{\hbar\omega{_\alpha}}{2}\Bigl[\hat p{_\alpha}^2 + \bigl(\hat q{_\alpha} -
g{_\alpha}\,\boldsymbol{\varepsilon}{_\alpha}\!\cdot\!\hat{\mathbf R}\bigr)^2\Bigr].
\end{eqnarray}
Eqs.~\eqref{eq:HL-qrep} and \eqref{eq:HL-prep} are related by a canonical $90^\circ$ rotation in the photon phase space, or equivalently by a global phase of the ladder operators; they are completely equivalent and appear in different papers as distinct {length-gauge} formulas.

\subsubsection{Diamagnetic term and dipole self-energy}
In the velocity gauge, expanding the kinetic energy produces the diamagnetic term
\begin{equation}
\label{eq:seagull}
\hat H{_{\mathrm{dia}}} = \sum{_i} \frac{{q_i}^2}{2{m_i}}\,\hat{\mathbf A}^2,
\end{equation}
often called the {``seagull''} term. It couples two photons to matter in a single vertex and is essential for gauge invariance, correct sum rules, and boundedness of the Hamiltonian.
Under the PZW transformation this term maps to the dipole self-energy in \eqref{eq:HL-decomposed},
\begin{equation}
\sum\alpha \frac{1}{2{\varepsilon_0} V}\bigl(\boldsymbol{\varepsilon}_\alpha\!\cdot\!\hat{\mathbf D}\bigr)^2,
\end{equation}
which plays the same stabilizing and gauge-enforcing role in the length gauge. Two-photon processes that are first-order via $\hat{\mathbf A}^2$ in the velocity gauge appear at second order in the linear $-\,\hat{\mathbf D}\!\cdot\!\hat{\mathbf E}$ coupling in the length gauge; the total physical predictions coincide.

\subsubsection{Differences of the two gauges in practice}
The two Hamiltonians are related by the exact unitary {$\hat H_L=\hat U^\dagger \hat H_v \hat U$} on the full light{--}matter Hilbert space, hence they have identical spectra and give identical results for all observables. Differences in which
bare Fock-manifolds couple (e.g., $\Delta n=\pm 1$ vs.\ $\Delta n=0,\pm 1,\pm 2$) are representation-dependent statements about the bare
photon basis, not physical differences. The ladder operators that diagonalize the free field after the PZW transformation are {``dressed''} by matter, e.g.,
\begin{equation}
\hat b{_\alpha} \equiv \hat U^\dagger \hat a{_\alpha} \hat U
= \hat a{_\alpha} + i\,\frac{g{_\alpha}}{\sqrt{2}}\,\boldsymbol{\varepsilon}{_\alpha}\!\cdot\!\hat{\mathbf R},
\end{equation}
so a {``}photon{''} in one gauge is a different superposition of light and matter in the other.

Exact equivalence requires working in the full Hilbert space and keeping $\hat{\mathbf A}^2$
(velocity gauge) or the dipole self-energy (length gauge). If one truncates the matter or photon subspaces
(few-level or few-photon models), naive truncations can break gauge equivalence.
Further practical differences between the two approaches arise from how
differential operators are discretized and the selection of basis
parameters (e.g., grid spacing in real space representation).


\providecommand{\latin}[1]{#1}
\makeatletter
\providecommand{\doi}
  {\begingroup\let\do\@makeother\dospecials
  \catcode`\{=1 \catcode`\}=2 \doi@aux}
\providecommand{\doi@aux}[1]{\endgroup\texttt{#1}}
\makeatother
\providecommand*\mcitethebibliography{\thebibliography}
\csname @ifundefined\endcsname{endmcitethebibliography}
  {\let\endmcitethebibliography\endthebibliography}{}

\end{document}